\documentclass[preprint]{aastex63}

\usepackage{epsfig,amsmath,amsfonts,amssymb,graphicx,subfigure,enumitem,url,natbib}

\received{........}
\revised{........}
\accepted{........}
\submitjournal{AJ}
\shorttitle{Polarization towards NGC 1817}
\shortauthors{Singh \& Pandey}

\begin{document}

\title{Broadband Linear Polarization in the Region of the Open Star Cluster NGC 1817}

\email{ssingh@aries.res.in} 
\email{jeewan@aries.res.in}

\author{Sadhana Singh}
\affiliation{Aryabhatta Research Institute of Observational Sciences (ARIES), Manora Peak, Nainital 263001, India}
\affiliation{School of Studies in Physics \& Astrophysics, Pt. Ravishankar Shukla University, Raipur-492010, India }
\author{Jeewan C. Pandey}
\affiliation{Aryabhatta Research Institute of Observational Sciences (ARIES), Manora Peak, Nainital 263001, India}

\begin{abstract}
Multi-band linear polarimetric observations of 125 stars in the region of the cluster NGC 1817 have been carried out  intending to study properties of interstellar dust and grains in that direction.  The polarization  is found to be wavelength-dependent, being maximum in the V band  with an average value of 0.95\%.  The foreground interstellar dust grains appear to be the main source of linear polarization of starlight toward the direction of NGC 1817. The average value of  the position angle in the  V band of 119\degr.2 is found to be less than the direction of the Galactic parallel in the region, indicating that the dust grains in the direction are probably not yet relaxed.  Spatial distribution of dust appears to be more diverse in the coronal region than the core region of the cluster.  The maximum value of the degree of polarization is estimated to be  0.93\% for members of the cluster using the Serkowski relation. The average value of wavelength corresponding to the maximum polarization of 0.54$\pm$0.02 $\mu$m  indicates that the size distribution of dust grains in the line of sight is similar to that of the general interstellar medium.  Several variable stars in the cluster were also observed for polarimetrically and pulsating variables appear to have a slightly lower value of polarization from other nonvariable member stars of the cluster. There are indications of the existence of dust layers existence in front of those clusters which are located close to galactic plane while for clusters located away from galactic plane no major dust layers are observed.
\end{abstract}   
   

\section{Introduction} \label{sec:intro}
Polarization of background starlight through the interstellar medium (ISM) is a result of a dichroic extinction of starlight through the aligned grains. The ISM dust grains are asymmetric and aligned in a direction by the magnetic field such that the short axis of grains is parallel to the magnetic field direction (also known as Davis-Greenstein type orientation). This well-accepted and popular mechanism was given by \citet{1951ApJ...114..206D} and  used in the modeling of ISM polarization and grain alignment. The polarimetric study will help us to achieve both macroscopic and microscopic information about the ISM.  By macroscopic properties  we mean the projected direction of the magnetic field at the region \citep{2007JQSRT.106..225L} whether it is a macroscopic field in the Galaxy \citep{1970MmRAS..74..139M,1976MNRAS.177..499A} or the local magnetic fields of the cloud \citep{1990ApJ...359..363G}, whereas the microscopic properties are the size distribution, shape, and efficiency of dust grains \citep{2008JQSRT.109.1527V}. Furthermore, polarimetry is useful in explaining the membership of the stars in the cluster \citep{2013MNRAS.430.1334M} and in detecting the location of energetic phenomena that happened in the history of a cluster \citep{2003ApJ...598..349F}. Thus the polarimetric technique is assumed as a more reliable technique to get the information about dust grains.

 Several open star clusters in the galactic  anticenter direction (i.e. $l$ $\sim$ 118\degr - 174\degr) have been studied in the past, e.g. Berkeley 59 \citep[$l$ = 118\degr.22; $b$ = 5\degr.0;][]{2012MNRAS.419.2587E}, Casado Alessi 1 \citep[$l$ = 123\degr.26; b = -13\degr.30;][]{2020AJ....159...99S}, NGC 654 \citep[$l$ = 129\degr.082; $b$ = -0\degr.357;][]{2008MNRAS.388..105M}, IC 1805 \citep[$l$ = 134\degr.726; $b$ = 0\degr.919;][]{2007MNRAS.378..881M},  NGC 1893 \citep[$l$ = 173\degr.585; $b$ = -1\degr.68;][]{2011MNRAS.411.1418E}, NGC 1931 \citep[$l$ = 173\degr.898; $b$ = 0\degr.281;][]{2013ApJ...764..172P} in which foreground dust grains are found to be responsible for polarization and a little fraction of stars had shown the indication of intrinsic polarization except for the stars in the cluster Berkeley 59 in which 40\% of cluster members were noted as having an intrinsic polarization component. The average distributions of size of the dust grains lying in front to these clusters were found to be similar to those for general ISM with the wavelength corresponding to the maximum polarization in the range of 0.52 - 0.59 $\mu m$. In some lines of sight,  more than one dust layer was  found to be responsible for the polarization, e.g., in the case of Berkeley 59, three dust layers at distance of 300, 500, and 700 pc were found to produce polarizations of $\approx$ 0.2-1.0, $\approx$ 1.0-3.0, and $\approx$ 5.5 \% with different magnetic field orientation, respectively. Whereas, for the clusters NGC 654, NGC 1893, and NGC 1931 two dust layers are found in front of them (see also the Section \ref{sec:diss}).  
Thus open star clusters are suitable candidates to observe as they spread within an area  so that the evolution of the dust parameters all over that region can be analyzed. Further, photometric and spectroscopic studies of many open clusters are easily available and by combining the polarimetric results with the available details one can extract other information like polarizing efficiency, foreground dust concentration, etc. 

Using the linear polarimetric observations, we would like to study properties of the  ISM dust and  grains toward the open star cluster NGC   1817 ($l$ = 186\degr.156, $b$ = -13\degr.096).
 This cluster lies almost directly toward the Galactic anticentre direction. The center of the cluster was determined as  R.A.(J2000)= 05$^{h}$12$^{m}$15$^{s}$ and decl.(J2000)= +16\degr41\arcmin 24\arcsec\ by  \cite{2009MNRAS.399.2146W}.  The cluster is an intermediate age open cluster with an age of 800 - 950  Myr,  distance of  1.8 - 2.1 kpc, and reddening ($E(B-V)$) of 0.27 - 0.33 mag  \citep{1977AJ.....82..612H, 1994A&AT....4..153L, 1997AJ....114.2556T, 2000A&A...359..347D,2004A&A...426..819B,2005A&A...438.1163K,2011AJ....142...59J}.
NGC 1817  is a relatively wide cluster with an angular radius of the core and cluster of 6\arcmin\, and 13\arcmin.8, respectively \citep{2005A&A...438.1163K}. More recently, using the Gaia data \citet{2018A&A...618A..93C} showed that half of the members of the cluster NGC 1817 lie within 11\arcmin.22 with a mean distance of 1.82 kpc.  Different types of variable and binary stars were also identified in the region of the cluster NGC 1817 and many of them are  members of the cluster \citep{2012A&A...548A..97Z}. More recently, \citet{2020AJ....159...96S} have presented a survey of variable stars in the cluster NGC 1817 using the  data obtained from the Kepler K2 Campaign.
A total of 44 red clump stars, 32 $\delta$ Scuti stars, 27 $\gamma$ Dor stars, 7 hybrid pulsators, and 13 eclipsing binaries have been identified in this cluster. Thus the cluster is also known to have the richest population of pulsating stars.

The paper is organized as follows. The observation, data reduction, and identification of stars are described in Section \ref{sec:obs}. Results and analysis from the polarimetric observations are given  in Section \ref{sec:result}. In Section \ref{sec:diss}, we discuss our results whereas a summary of our findings is  given in the Section \ref{sec:sum}.

\section{Observations, data reduction, and identification}\label{sec:obs}
We have observed the cluster NGC 1817  by using the Aries IMaging POLarimeter (AIMPOL) on 2018 January 18, 19, and 22  in four passbands B, V, R, and I. The AIMPOL is mounted as a backend of 104 cm Cassegrain telescope of Aryabhatta Research Institute of Observational Sciences (ARIES). AIMPOL consists of a half-wave plate (HWP) and a Wollaston prism, and as a result two images, one ordinary and another extraordinary, are formed for a single source on the detector chip. The detector is a Tek 1k$\times$1k charge-coupled camera with a gain of 11.98 e$^{-}$/ADU and readout noise of 7.0 e$^{-}$. The total field of view of AIMPOL is 8\arcmin\, in diameter.  A detailed description of AIMPOL is given in \cite{2004BASI...32..159R}. To observe the entire field of the cluster NGC 1817, we have divided it into six regions for observations.  Stars HD 21447 and HD 25443 were observed as standard unpolarized and polarized stars in all four bands for applying the correction for instrumental polarization and position angle, respectively. Observations were taken at four positions of HWP  of 0\degr, 22\degr.5, 45\degr, and 67\degr.5 from the north-south direction with at least three frames at each position. The exposure times were in the range of 60-250 s in different bands. To increase the signal-to-noise ratio, we have summed all three frames taken at each position of the HWP. In order to obtain the fluxes of ordinary and extraordinary images of each star in the cluster, we have performed the aperture photometry using Image Reduction and Analysis Facility\footnote{http://iraf.net}. Details of the reduction techniques are given in \citet{2020AJ....159...99S}.  A total of 125 stars in the region of NGC 1817 were selected to derive the polarization values. The selection is biased by two factors: an observation limit of $V$ $\sim$ 16 mag and two images of a star should be well isolated from the other two images of neighboring stars. Finally,  the isolation of each stellar image was also checked manually.
The instrumental polarization was found to be  $\leq$ 0.3 \% in all bands as also found in earlier studies \citep[e.g.][]{2004BASI...32..159R,2009MNRAS.396.1004P,2013MNRAS.430.2154P, 2016MNRAS.457.3178P,2020AJ....159...99S}. By applying a correction to the instrumental polarization the results for polarized standard stars were found to be consistent with standard values as given in \citet{1992AJ....104.1563S}.
Both the degrees of polarization and position angles of all 125 stars in the field of NGC 1817 were corrected for instrumental polarization and zero-point correction of the  position angle. 
 
The astrometry was performed using the online available tool\footnote{https://nova.astrometry.net/upload} to get the sky position of the observed stars in the field of NGC 1817. For the identification of these 125 observed stars, we have made use of the Gaia archive.  After applying the criteria of the  five-parameters solution \citep[see][]{2018A&A...616A...2L}, we have extracted the positions and parallax of all the sources in the 12\arcmin\, field of NGC 1817 from the Gaia DR2 data \citep{2016A&A...595A...1G,2018A&A...616A...1G}. 
All 125 polarimetrically observed stars were then cross-matched with the Gaia source positions using the CDS X-Match Service\footnote{http://cdsxmatch.u-strasbg.fr/}, by applying position cross-match criteria in the all sky area.   All the observed sources were matched with Gaia sources within  1\arcsec.2. In Table \ref{tab:pthe}, we give our assigned ID, Gaia ID, and the offsets from the Gaia positions of all 125 cross-matched stars along with their polarization values in all four bands.  We have adopted membership information for the stars in the cluster NGC 1817 from the catalog of \cite{2018A&A...618A..93C} due to the precise astrometric information from the Gaia data. Among 125 observed sources, 68 stars are found with  a membership probability $>$ 50 \% from \citet{2018A&A...618A..93C}. For further analysis, we have taken these stars as members and others as nonmembers. The member stars of the cluster NGC 1817 are marked with the asterisk symbol with their ID. Out of 125 stars, only one star (ID 6) was found to have a parallax value under 2$\sigma$ in the Gaia archive, and thus was not used for any further analysis where distance is required. The 50$^{th}$ percentile value of extinction $A_{V}$ for different stars in NGC 1817  were taken from \citet{2019A&A...628A..94A} whereas the value of color excess [$E(B-V)$] for each star is calculated using the normal reddening law $R_{V}$ = 3.1.

\section{Analysis and Results} \label{sec:result}
\begin{figure*}
	\centering
\includegraphics[width=0.80\columnwidth]{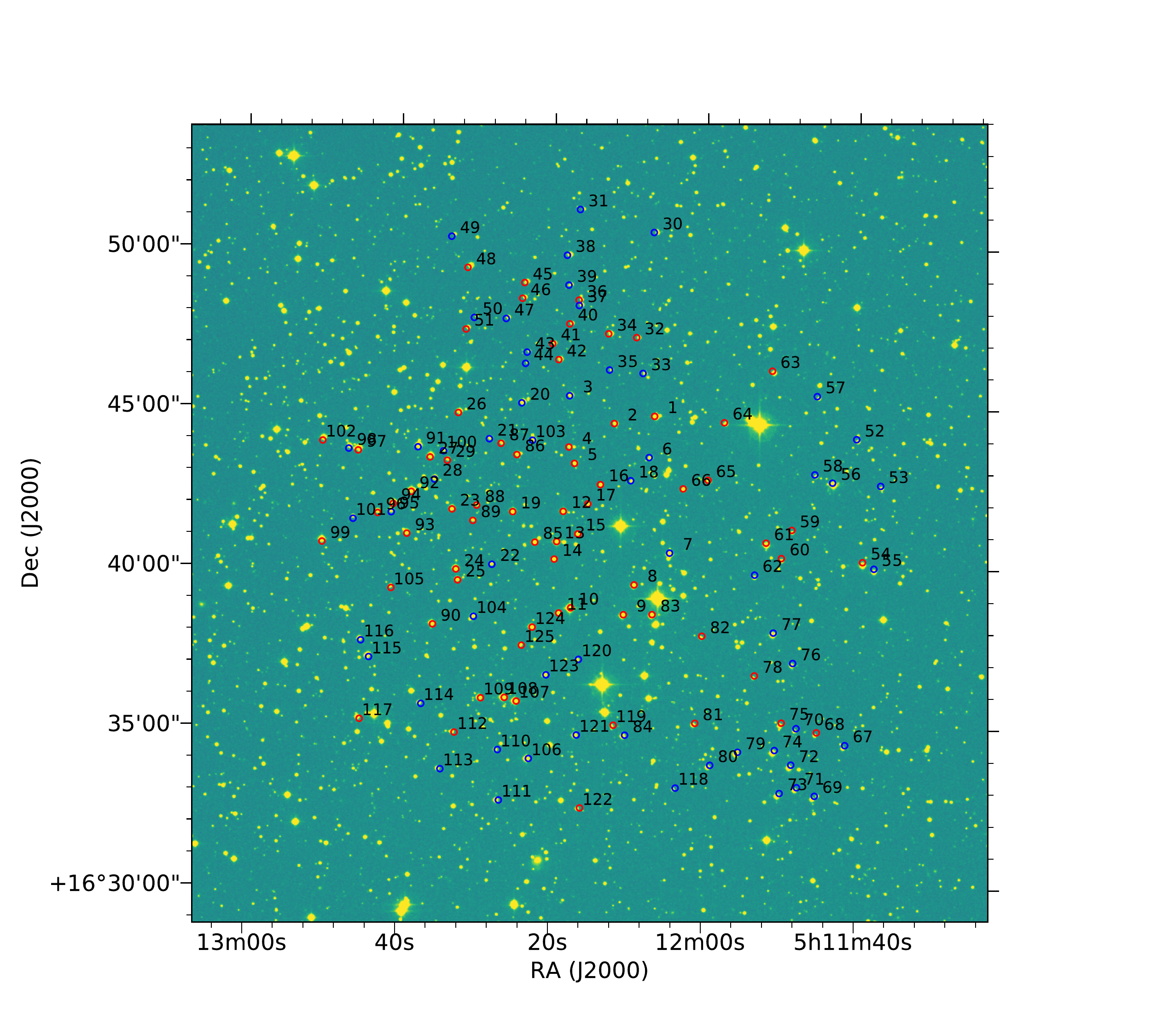}
\caption{Position of observed stars in the sky plane. Red open circles are for members and nonmembers are denoted by blue open circles. }
    \label{fig:coord}
\end{figure*}

The degrees of polarization ($P$) and position angles ($\theta$) in B, V, R, and I bands for all 125 stars observed in the field of NGC 1817 are given in Table \ref{tab:pthe}. The position of all these stars along with the star ID as assigned in Table \ref{tab:pthe} in the sky is shown in Figure \ref{fig:coord}. The red open circle denotes the member stars whereas the blue open circles denote nonmembers. The observed member stars are concentrated mainly in the center while nonmembers are located in the outskirts, mainly toward the south and northeast part of the region. Very few nonmembers are located in the center region.

\subsection{Sky Projection of Polarization Vectors and the  Distribution of $P$ and $\theta$} \label{dss}
\begin{figure*}
	\centering
\subfigure[Polarization vectors overlaid on DSS image.]{\includegraphics[width=0.45\columnwidth]{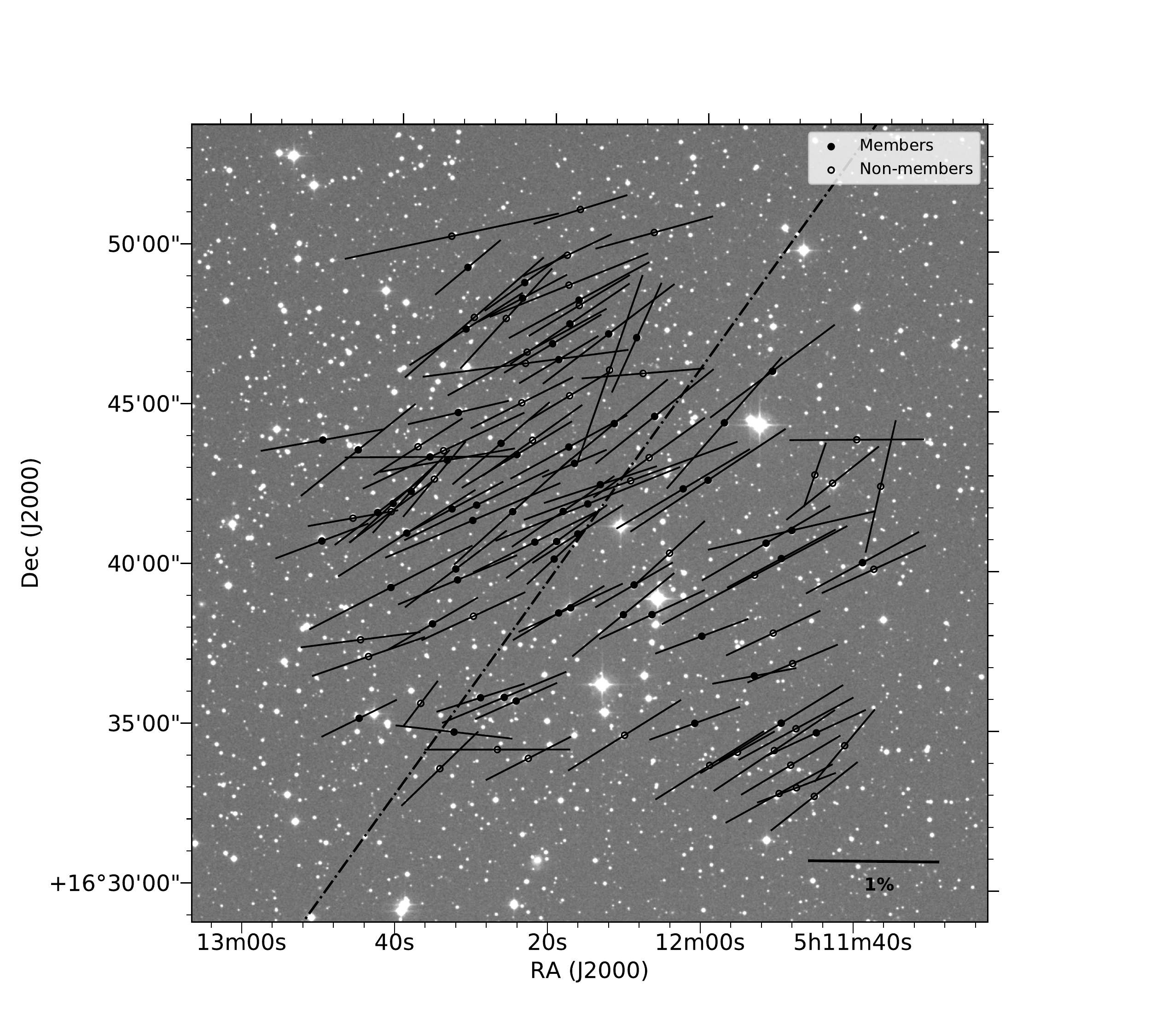}\label{fig:dss_vector}}
\subfigure[Histogram of $P_V$.]{\includegraphics[width=0.40\columnwidth]{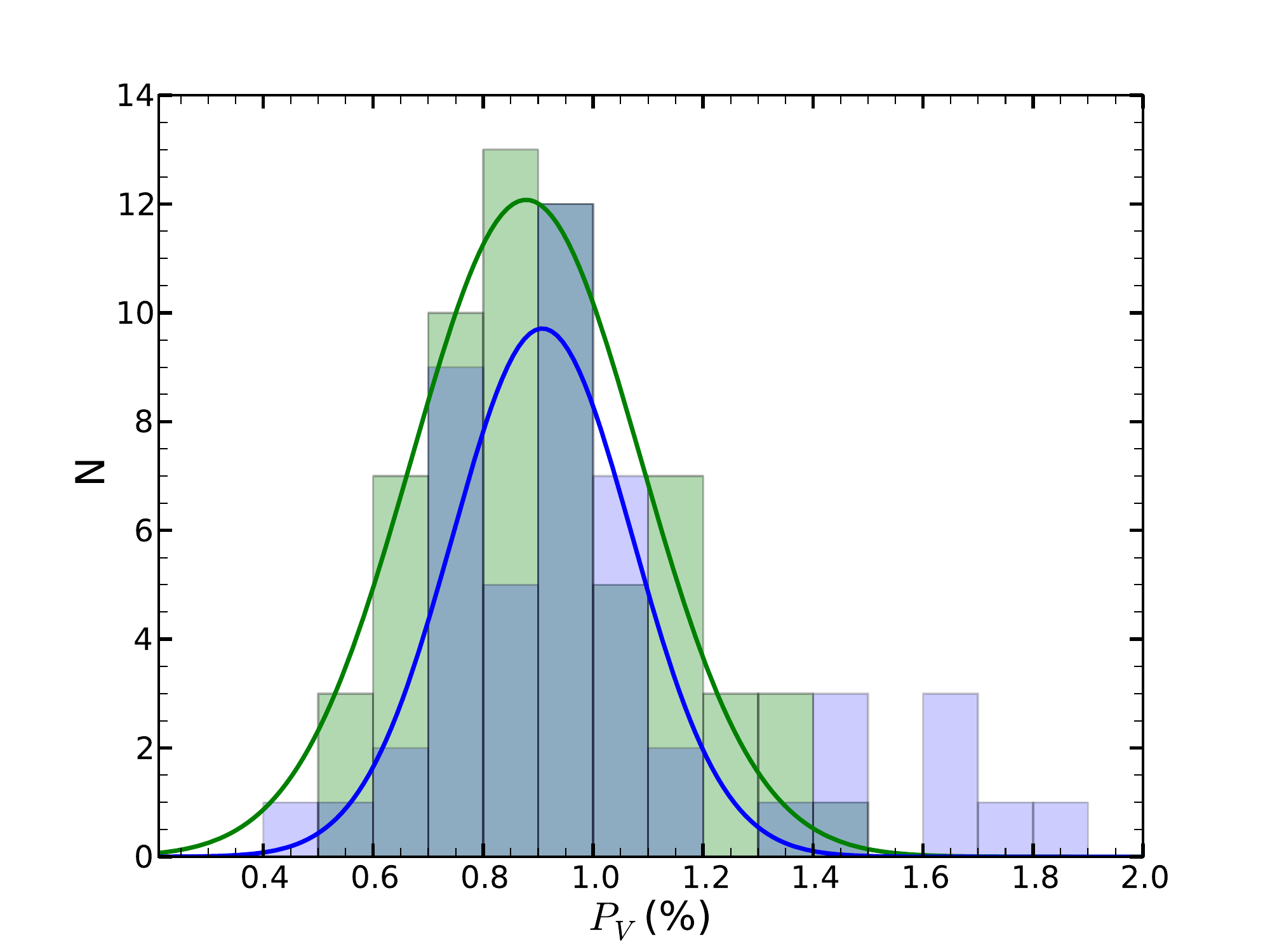}\label{fig:pv_hist}} 
\subfigure[Histogram of $\theta_V$.]{\includegraphics[width=0.40\columnwidth]{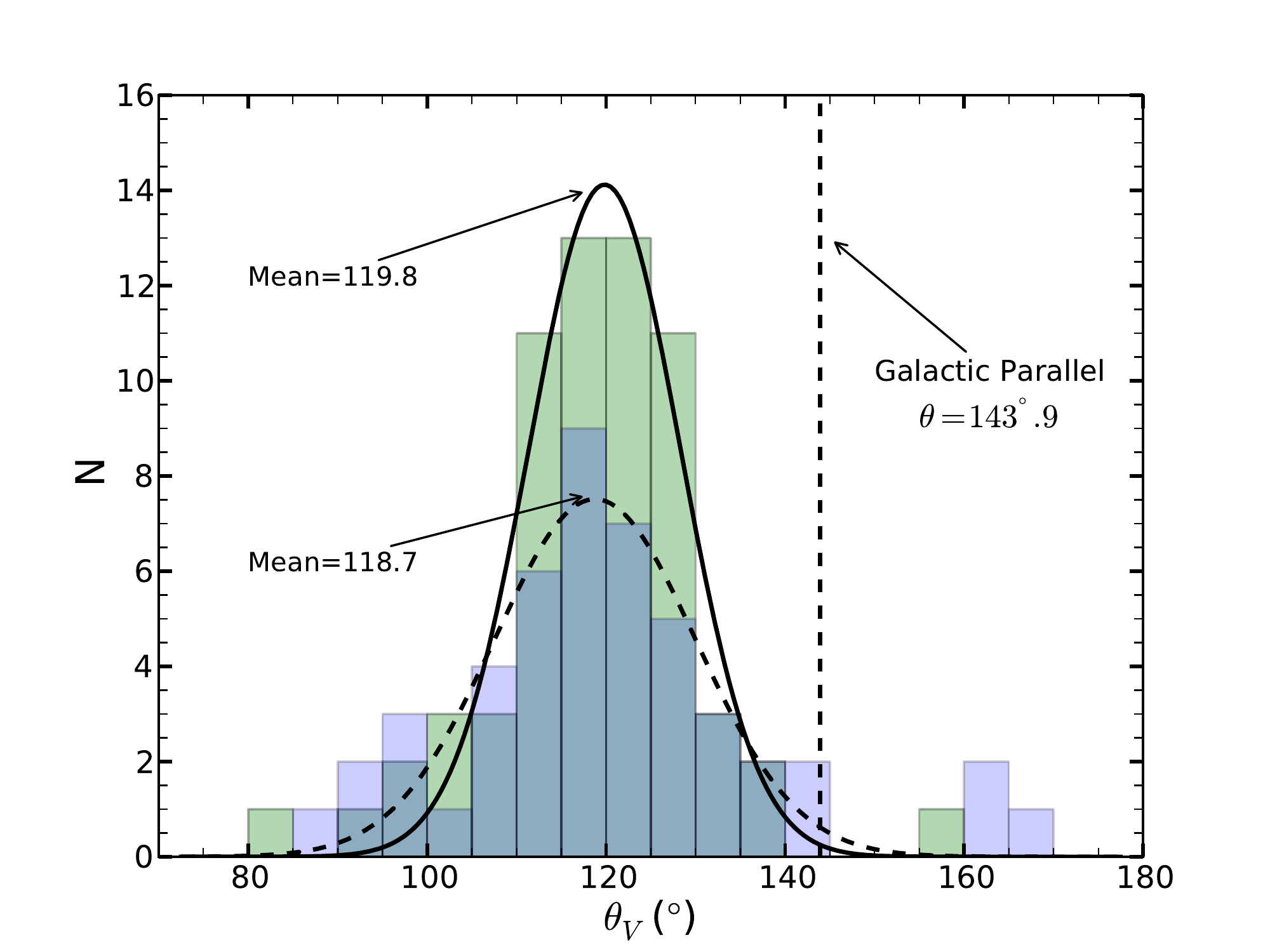}\label{fig:thv_hist}}
\subfigure[$P_V$ versus $\theta_V$.]{\includegraphics[width=0.40\columnwidth]{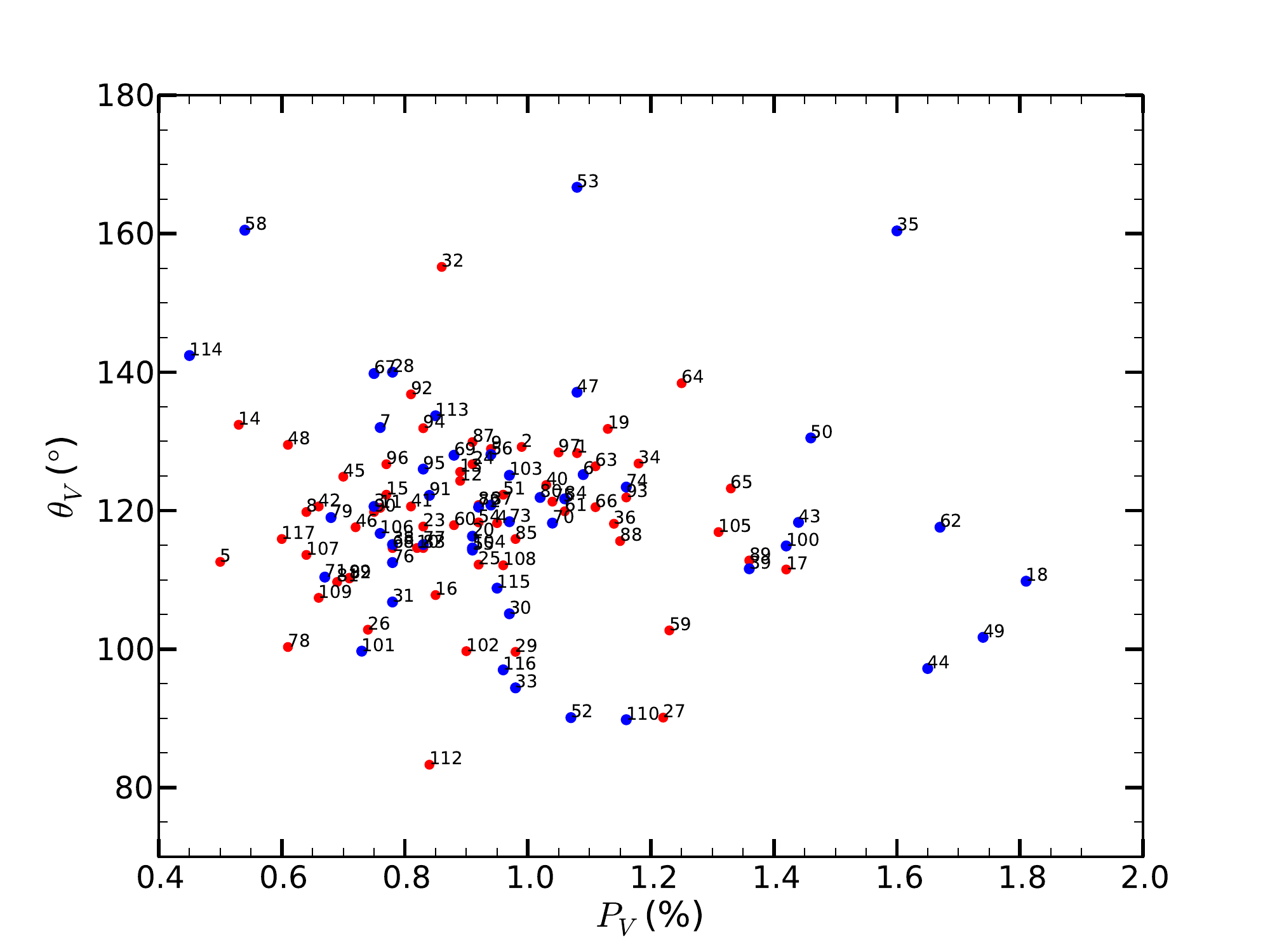}\label{fig:pv_thev}}
\subfigure[Spatial distribution of $P_{V}$.] {\includegraphics[width=0.40\columnwidth]{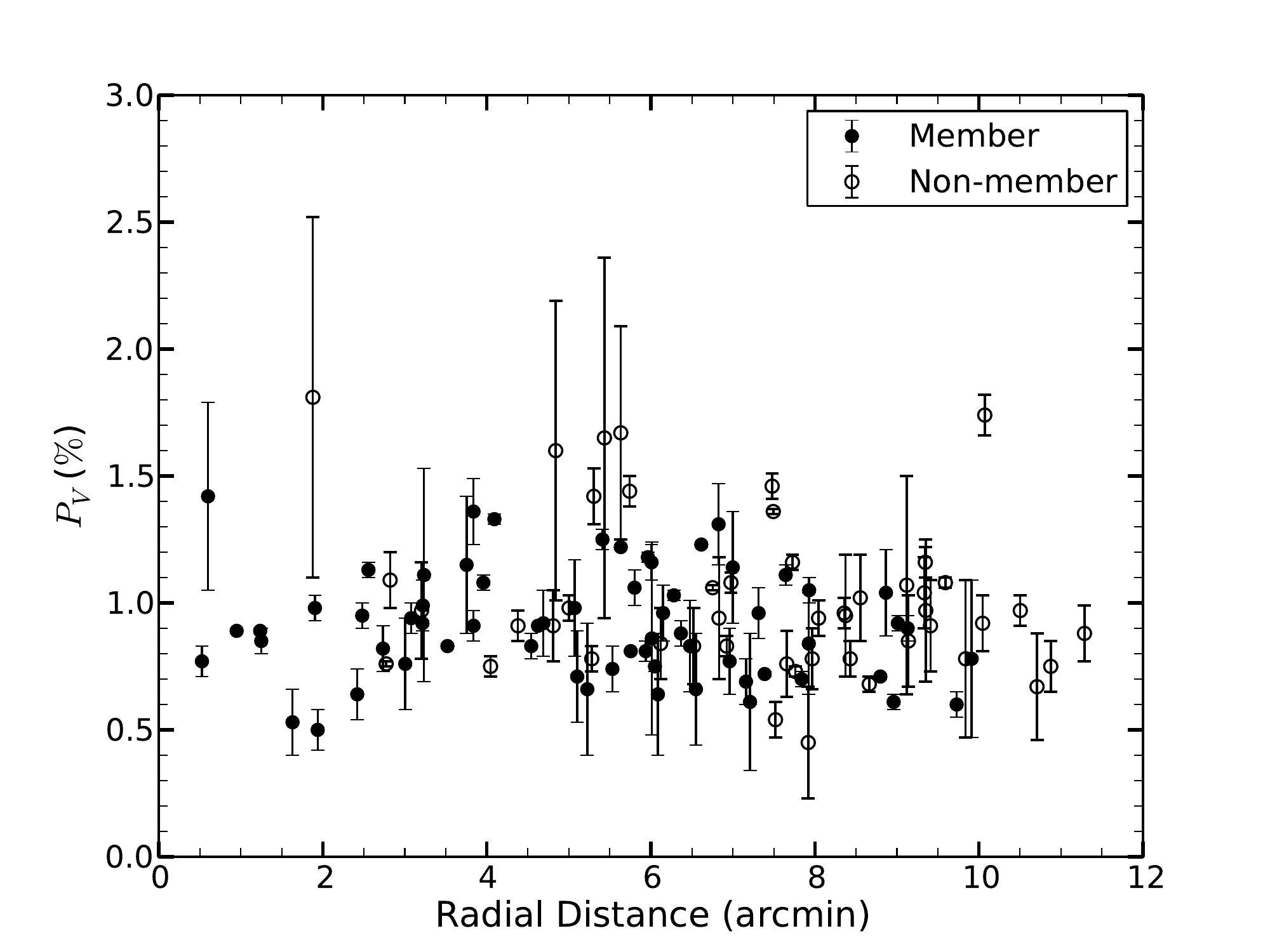}\label{fig:pv_spatial}}  
\subfigure[Spatial distribution of $\theta_{V}$.] {\includegraphics[width=0.40\columnwidth]{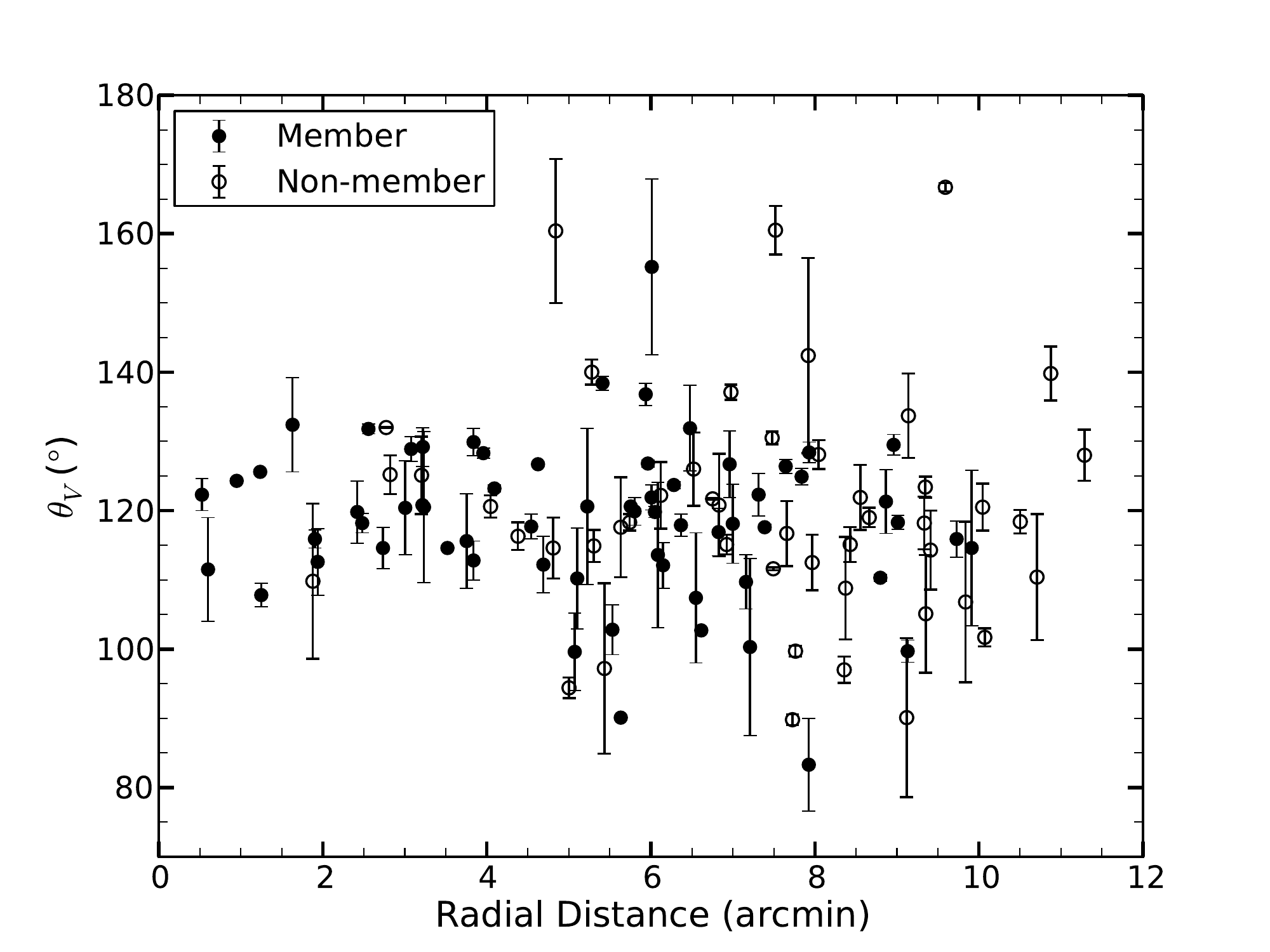}\label{fig:tv_spatial}}

\caption{(a) Polarization vectors for the observed stars in the V band are overplotted in the sky plane. The stars with the filled circle symbols are members and open circle symbols are for nonmembers. The length of the vector signifies the degree of polarization and the reference for 1\% polarization is shown at the bottom right. The position angles are measured from the north increasing toward the east. The dashed-dotted line specifies the projection of the orientation of the GP at the region. (b), (c) Distribution of $P_{V}$ and $\theta_V$ for  members and nonmembers along with the Gaussian fitting. Members are shown in green and nonmembers are in blue. The dashed vertical line in Figure  \ref{fig:thv_hist} shows the projection of the GP in the region. (d)  In the $P_V$ versus  $\theta_V$ diagram, red and blue dots represent for members and nonmembers, respectively. IDs of all the stars, as assigned in the Table \ref{tab:pthe}, are also mentioned along with the corresponding markers. (e), (f) The variation of $P_{V}$ and $\theta_{V}$ with the radial distance from the cluster center.}
    \label{fig:ptheta}
\end{figure*}

The sky projection of polarization vectors for the stars in the V band over the digitized sky survey\footnote{http://archive.eso.org/dss/dss} image of the cluster is shown in Figure \ref{fig:dss_vector}. The length of the  vectors denotes the degree of polarization and their orientation is the polarization position angle from the north increasing toward the east. A reference polarization vector for 1 \% polarization is shown at the bottom right of the figure. Member stars are shown by filled circles whereas the open circle notation is for nonmembers. A dashed-dotted line indicates the orientation of the projection of the Galactic plane (GP) in this region. The polarization vectors were not found to be parallel with the direction of the GP.

The values of $P$ were found to be in the ranges of 0.23 \% - 2.33 \%, 0.45 \% - 1.81 \%, 0.44 \% - 2.91 \%, and 0.12 \% - 1.34 \% in the B, V, R, and I bands, respectively (see Table \ref{tab:pthe}). The $\theta$ for the majority of stars lies nearly at the same value in all bands. The $\theta$ extent is from 65\degr.6 to 179\degr.9, 83\degr.3 to 166\degr.7, 80\degr.6 to 177\degr.7, and 67\degr.0 to 178\degr.9 degree in the B, V, R, and I bands, respectively.
The average values of $P$ and $\theta$  of all observed stars, members, and nonmembers in the B, V, R, and I bands are given in Table \ref{tab:ptheta}. The maximum value of the average polarization was found near to the V band and decreased toward the B and I bands. The distributions of $P_{V}$ and $\theta_{V}$ are given in the Figures \ref{fig:pv_hist} and \ref{fig:thv_hist} for both member and nonmember stars along with the best-fit Gaussian curve. A spread of $\Delta P_{V}$ = 0.92 \% is observed for member stars. It was found that the distribution of members and nonmembers are overlapping over each other.  A similar trend was also seen in other bands. The dotted straight line at $\theta$ =143\degr.9 in Figure \ref{fig:thv_hist} denotes the angle for the GP.  The GP was found to deviate by an angle of around 25,  which is similar to that seen in Figure \ref{fig:dss_vector}.

The relation between $P_{V}$ and $\theta_{V}$ for all stars is shown in Figure \ref{fig:pv_thev}. Red and blue circles are for members and nonmembers, respectively, with their corresponding IDs. The position angle  $\theta_V$ appeared to increase with  $P_{V}$ up to a value of $P_{V} \sim$ 1\%  after that a slight decrease in $\theta_{V}$ was noticed. This pattern is much clearer for member stars of NGC 1817.  A very dense group of stars is seen from $P_{V}$ = 0.6 \% to 1.2 \%.  However, after $P_{V}$ = 1.2\%, a scattered pattern  is clearly seen. One member star (ID 32) was found to have a very large position angle ($\theta_{V}$ = 155\degr.2) in comparison to the other members whereas another two member stars (ID: 27 and 112) have small $\theta_{V}$ (i.e. $<$ 95\degr). However, in nonmembers, four stars (ID: 114, 58, 53, and 35) and two stars (ID 52, 110) are found with $\theta_{V}$ more than 140\degr and less than 95\degr, respectively.  All the blue stragglers were found in a closed/compact group in this figure except for one blue straggler (ID 19) which has a higher value of  $P_{V}$ (= 1.13 \%) than other blue stragglers. 

The distribution of polarization parameters ($P_{V}$ and $\theta_{V}$) with radial distance from the center are shown in Figures \ref{fig:pv_spatial} and \ref{fig:tv_spatial}, respectively. The coordinates for the center of the cluster NGC 1817 are taken as 05$^{h}$12$^{m}$15$^{s}$, +16\degr41\arcmin 24\arcsec\, \citep{2009MNRAS.399.2146W}. Member stars are shown by filled circles and nonmembers by open circles.
The distributions of both  $P_{V}$  and $\theta_V$ were found to be almost constant up to a cluster radius of $\sim$ 5\arcmin, after that a scattered distribution was seen for both members and nonmembers.

\subsection{Polarization versus Distance} \label{dust_layer}
\begin{figure*}
	\centering
\subfigure[The polarization in V band as a function of distance.]{\includegraphics[width=0.48\columnwidth]{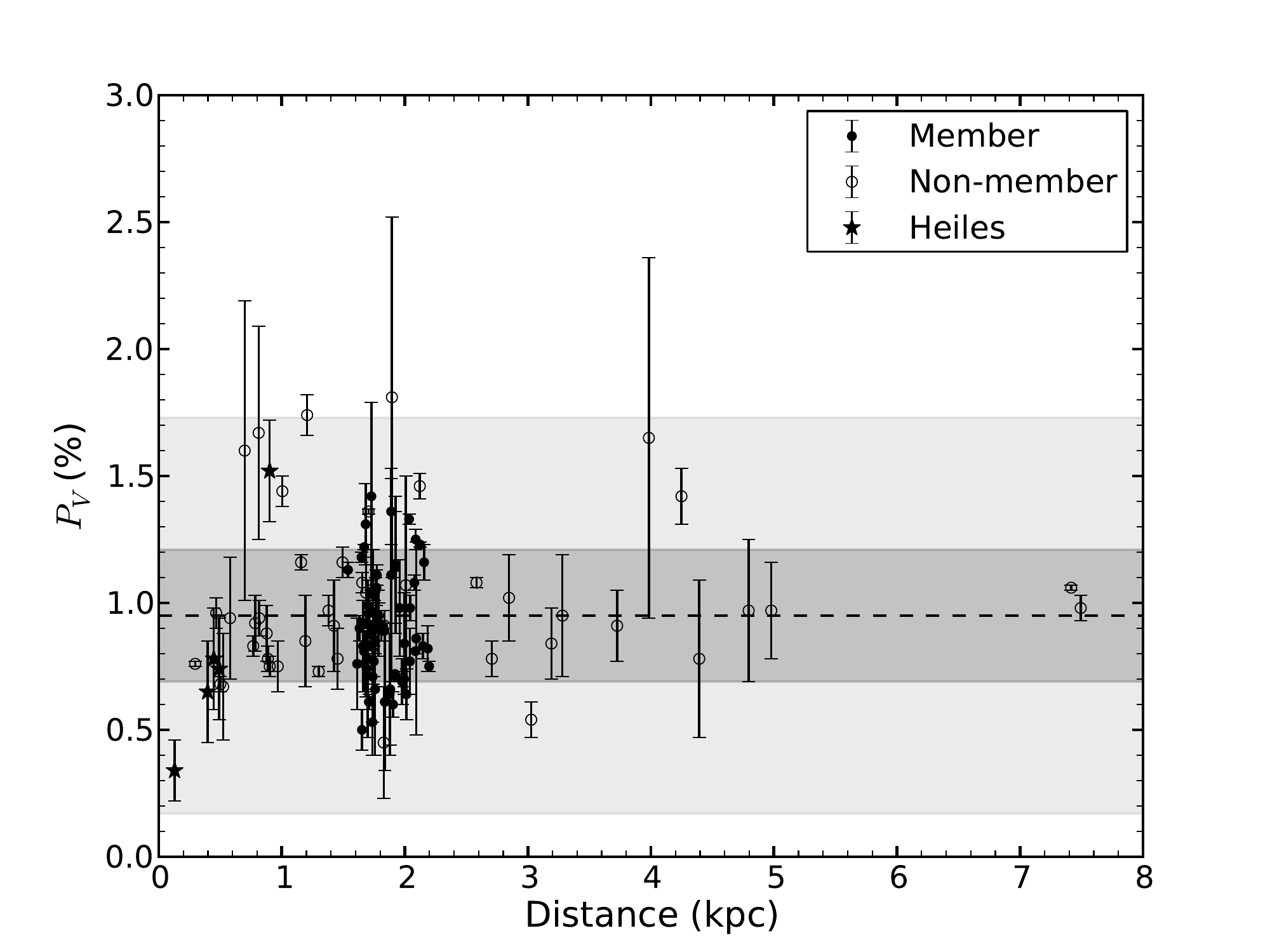} \label{fig:p_dist}} 
\subfigure[The variation of extinction with distance.]{\includegraphics[width=0.48\columnwidth]{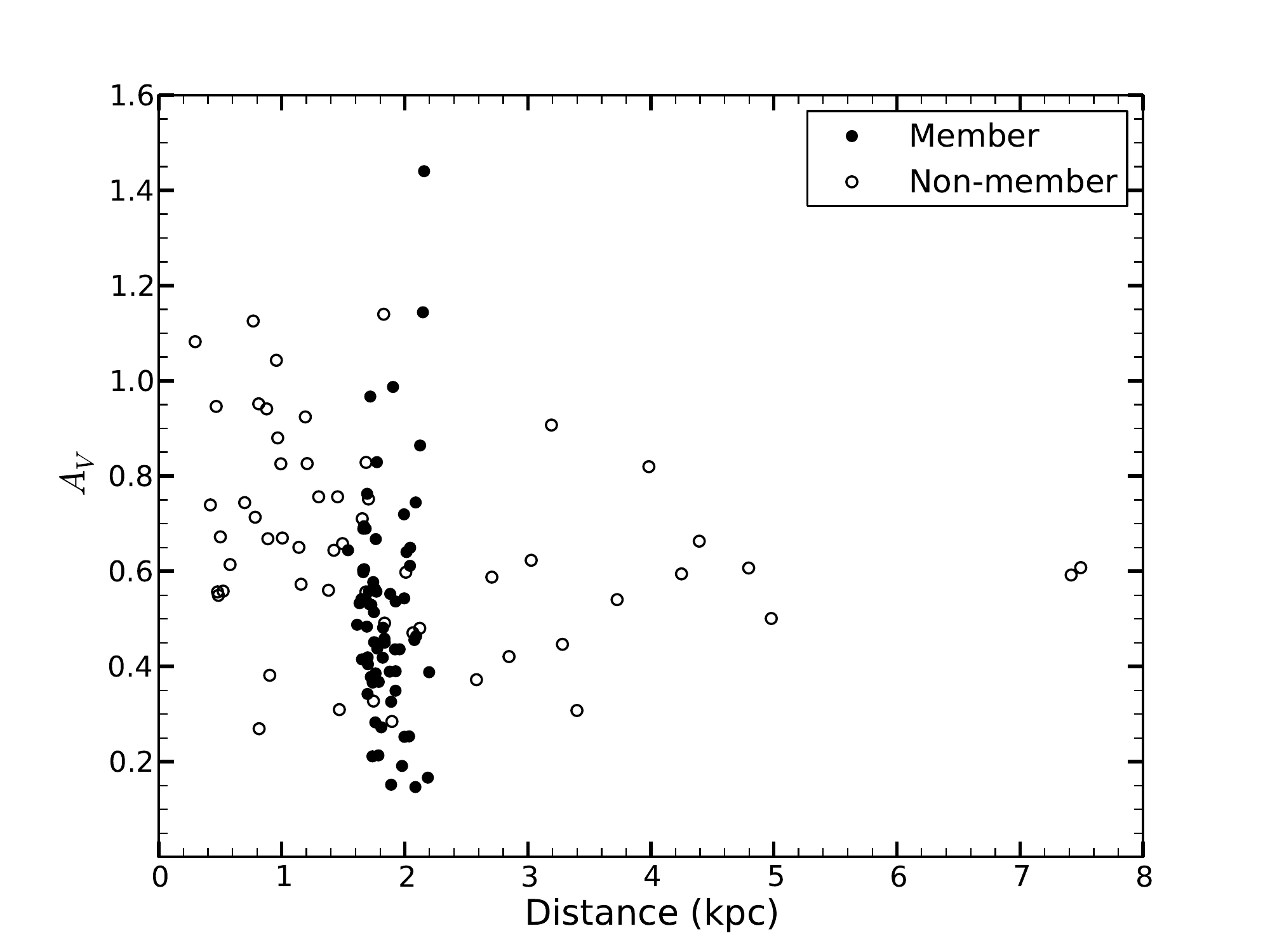} \label{fig:Av_dist}} 
    \caption{The polarization and extinction are plotted with the distance. Members are shown by filled circles and nonmembers by open circles. (a) The dotted horizontal line denotes the mean value of $P_{V}$. The dark and light gray shaded areas correspond to $\pm$1$\sigma$ and $\pm$3$\sigma$ regions around the mean value.}
    \label{fig:pAv_dist}
\end{figure*}

The variation of polarization with distance is crucial for understanding the distribution of interstellar matter. In the line of sight, between observer and the source, usually the dust  is located in layers and as a result the polarization is not a continuous function of the distance. If the light encounters at a dust component at a particular distance then the trend of polarization will show a sudden jump at that distance \citep{2008MNRAS.388..105M,2010MNRAS.403.1577M,2012MNRAS.419.2587E}. Hence, the number of sudden jumps in the value of $P$ along the distance will indicate the number of dust layers encountered in that particular line of sight. With this aim, we have plotted $P_V$ as a function of distance in Figure \ref{fig:p_dist}. Here, the members and nonmembers are shown by filled circles and open circles, respectively. The dark and light gray shaded regions around the mean value of $P_{V}$ are $\pm 1\sigma$ and $\pm 3\sigma$, respectively.  We have also taken $P_{V}$ from the catalog of  \citet{2000AJ....119..923H}.  In the search radius of 5\degr\, from the cluster center only seven stars are listed in this catalog, out of which five stars are used in our study as the  other two stars have unreliable polarization values due to large errors. The distances of these stars were also taken from Gaia DR2. In Figure \ref{fig:p_dist}, these stars are marked by the star symbol. All members of the cluster NGC 1817 are clustered around  a distance of 1.8 kpc. Observed nonmembers lie at a distance of 0.3 to 7.5 kpc.  The $P_V$ was found almost constant along the distance.
As all types of dust grains contribute to extinction, so the occurrence of the dust component can be inferred by the plot of extinction versus distance as well \citep[see also][]{2010MNRAS.403.2041V}. Therefore, we have also plotted the extinction ($A_{V}$) as a function of distance in Figure \ref{fig:Av_dist}. The scattered distribution of $A_{V}$ with distance is noticed.

\subsection{Stokes Plane} \label{qu}
\begin{figure*}
	\centering
\includegraphics[width=0.79\columnwidth]{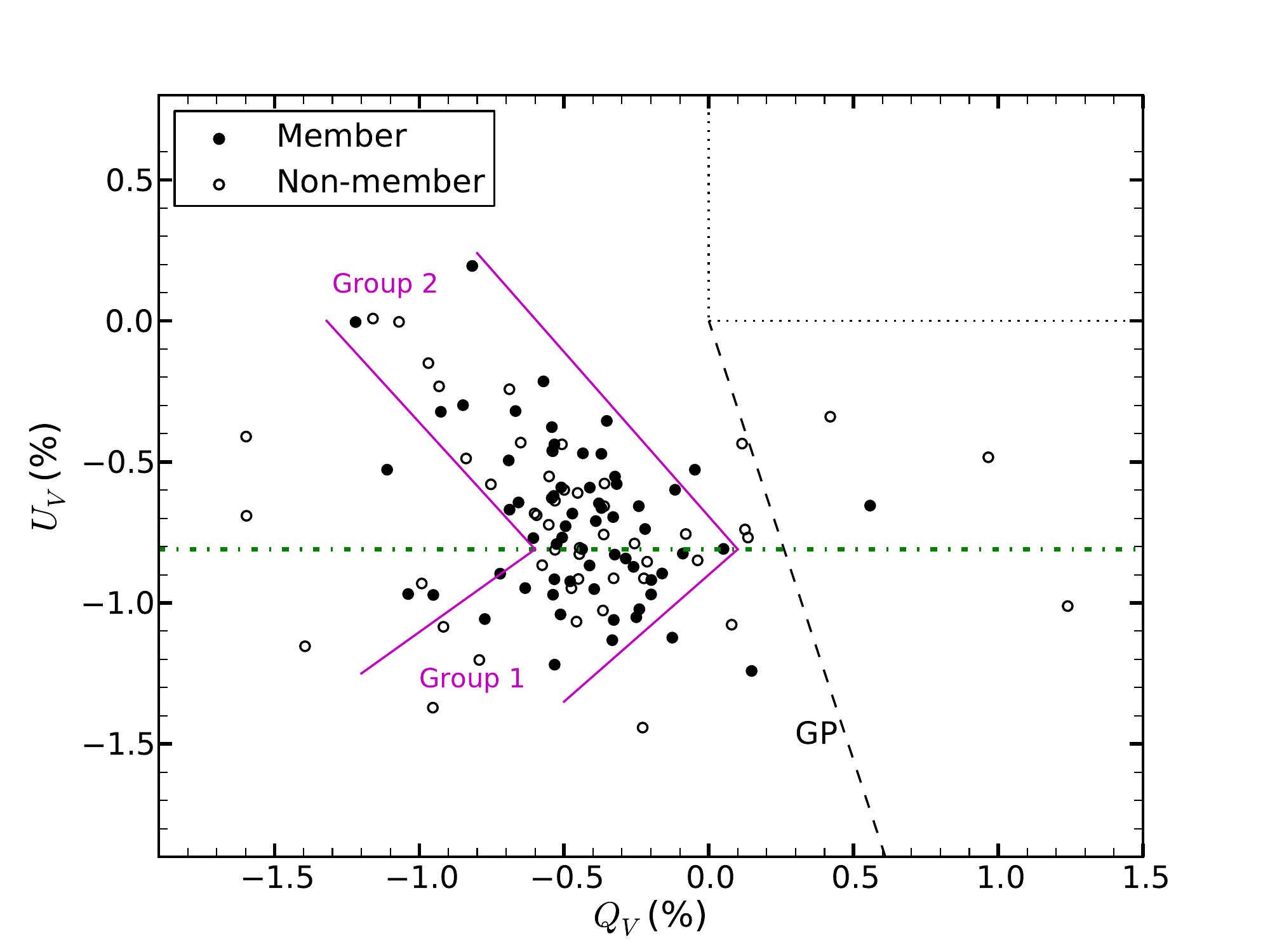} 
\caption{A plot between Stokes parameters $Q_{V}$ and $U_{V}$ with the direction of the GP as a dashed line. The lines in magenta color are drawn to show the patterns observed. The boundary at $U_{V}$=-0.81\% is marked by a green dashed-dotted line whereas the ($Q_V$,$U_V$) =(0,0) is the dustless solar neighborhood. }\label{fig:QU}  
\end{figure*}

The evolution of the interstellar environments from the Sun to the cluster can be also investigated by plotting Stokes parameters in the $Q$ (= $P cos 2\theta$) -  $U$ (= $P sin 2\theta$) plane. The ISM polarization is considered to be due to the column density of dust grains along the line of sight.
The members of the cluster are at same distance hence light suffers the same column density for all the stars of the cluster, and as a result the representative points of the members in the $Q$ - $U$ plane should concentrate in a region. Whereas for the foreground or background stars the column density differs so they may show a scattering pattern in the $Q$ - $U$ plane.  Besides this, if there are dust clouds/layers in between the observer and the cluster region, then the stars lying before and behind the clouds will show different values of $P$ and $\theta$ based on the polarimetric characteristics of the underlying cloud. A plot between $Q_{V}$ and $U_{V}$ is shown in Figure \ref{fig:QU}, where  members and nonmembers are shown by the filled and open circles, respectively. The position ($Q_V$,$U_V$) = (0,0) represents the dustless solar neighborhood and dashed line shows the direction of the GP.

We have noticed two patterns for both members and nonmembers as shown by group 1 and group 2 in the Stokes plane. Member stars falling in group 1 have an average value of ($Q_V$, $U_V$) of (-0.3514, -0.9429) whereas in group 2 member stars have an average value of (-0.4854,-0.5915). To discriminate both groups, we have plotted a dotted line along with $U_V$ = -0.81\%. This boundary was selected by observing the pattern and histogram of  $U_{V}$ below and above the value -0.81\%. In the first group a total of 25 member stars are present and in group 2 a total of  39 member stars are present. The average values of $\theta_V$ for group 1 and group 2 are described as 125\degr\, and 115\degr, respectively. It is apparent that group 2  is farther away from the GP than group 1.

\subsection{Serkowski Relation} \label{ser}
The ISM-originated polarization shows the maximum value of $P$ near the visual band and monotonically decreases toward the ultraviolet and infrared bands. A similar trend was found in the polarization value for the majority of the stars in the cluster NGC 1817. Also, the average value of $P$ was found as being the maximum in the V band (see Table \ref{tab:ptheta}).  The wavelength-dependent polarization can be used to compute the distribution of grain sizes on which the interstellar extinction law is dependent. The intrinsic polarization of a star due to either a circumstellar shell or any other asymmetry may show an abnormal wavelength-dependent polarization. \citet{1975ApJ...196..261S}  observed the wavelength dependence of interstellar linear polarization and gave an empirical relation between the degree of polarization $P_{\lambda}$ and  $\lambda$ as

\begin{equation}
P_{\lambda}/P_{max} =  exp[-1.15 \times ln^2(\lambda_{max}/\lambda)]
\label{eq:serko}
\end{equation} 

\noindent
Here, $P_{max}$ is the maximum polarization and $\lambda_{max}$ is the wavelength corresponding to the maximum polarization which depends on the optical properties and characteristic particle size distribution of the aligned grains \citep{1978ApJ...225..880M,1980ApJ...235..905W}. We have fitted the Serkowski relation as stated in the equation \ref{eq:serko} to the $P_{\lambda}$ versus $\lambda$ for all observed stars excluding those for which polarization measurements were only for two bands.  The values of $P_{max}$, $\lambda_{max}$, and $\sigma_{1}$ for these stars with their IDs are given in Table \ref{tab:serkowski}. Members are marked with the asterisk (*) symbol with the corresponding  IDs.  Stars for which the fitted parameters were obtained with $>$ 2 $\sigma$ are not included in the Table \ref{tab:serkowski}.  
The ISM-induced polarization can be known by knowing the best-fit parameter  $\lambda_{max}$  and/or the parameter $\sigma_{1}$, the unit weight error of fit. If the value of $\sigma_{1}$ is less than  1.6 and/or the value of $\lambda_{max}$ lies in the range of 0.45 to 0.80 $\mu m$, then polarization can be considered to be ISM originated. Otherwise, polarization is considered to be intrinsic \citep{1975ApJ...196..261S,1998AJ....116..266O}.  The Serkowski relation could not fit well due to the higher values of $\sigma_{1}$( i.e. $>$ 1.6) for three member stars (IDs 13, 27, and 40). Star 13 is a giant star, star 27 is a spectroscopic binary (SB),  and star 40 is a pulsating star. However, for star 40, the polarimetric observations were carried out only in three passbands. For eight of the  member stars (ID: 4, 9, 24, 32, 54, 65, 68, and 97)  low values  of $\lambda_{max}$ ($<$0.4 $\mu m$) were obtained.  Out of these eight stars, for four stars (ID 54, 65, 68, and  97) a large value of $P_{max}$  was also obtained.  For the three nonmember stars (ID 18, 58, and 70) the Serkowski relation fit well  with a $\sigma_1$ value $< 1.6$ but we have obtained  smaller  values of $\lambda_{max}$ and higher $P_{max}$. 

\begin{figure*}
\centering 
\includegraphics[width=0.69\columnwidth]{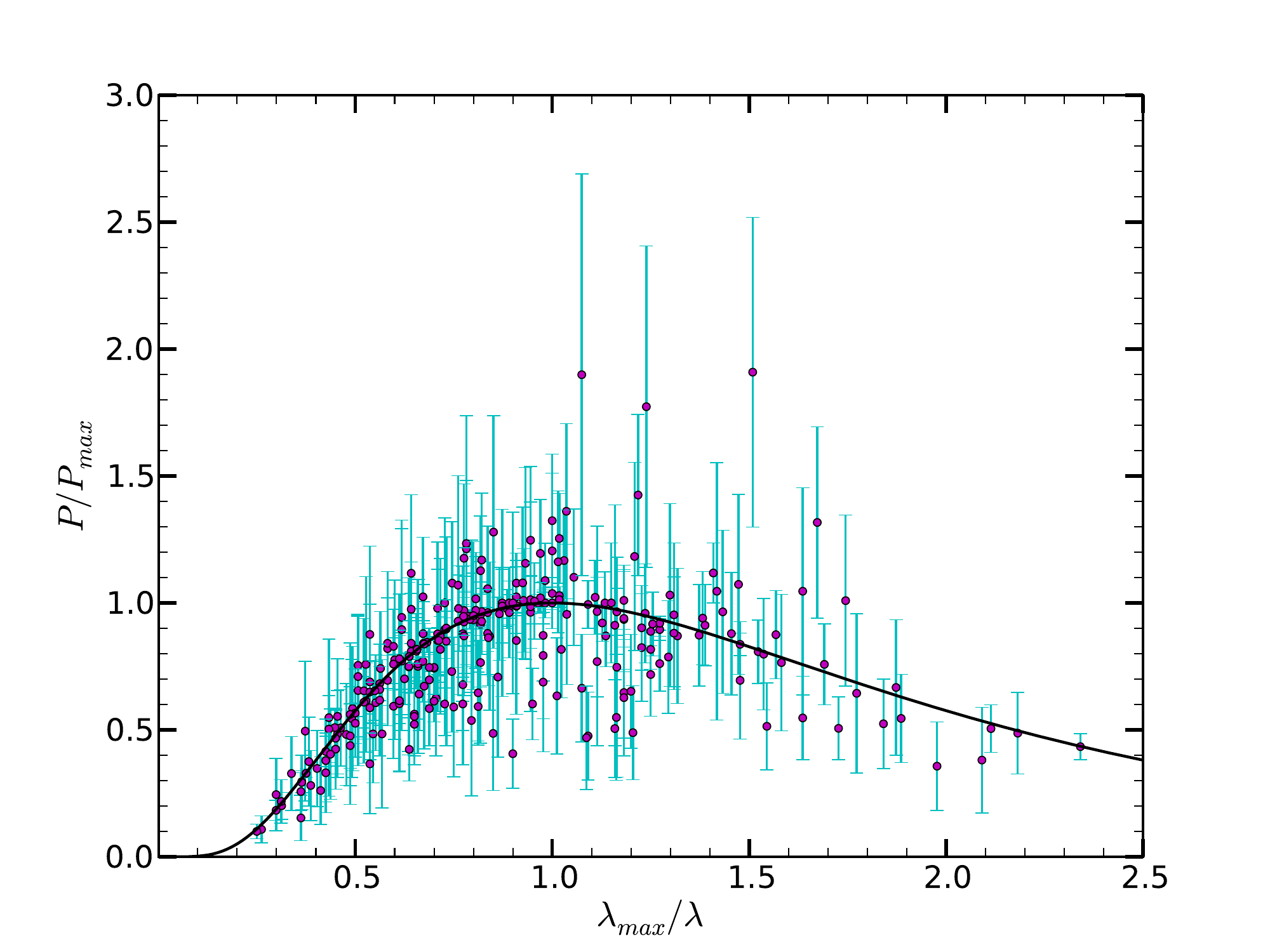}
    \caption{Normalized polarization wavelength-dependence plot for the observed stars. The black curve denotes the Serkowski relation for the general ISM.}
    \label{fig:serkow}
\end{figure*}

The average values of $P_{max}$ and $\lambda_{max}$ were found to be 0.93$\pm$0.03 \% and 0.54$\pm$0.02 $\mu m$, respectively for the cluster members after excluding stars that show an indication of intrinsic polarization. The plot between $\lambda_{max}/\lambda$ and $P/P_{max}$ is shown in Figure \ref{fig:serkow}.  The curve denotes the Serkowski empirical relation for the general ISM. The similarity between observations and the curve indicates that the polarization is dominated by ISM polarization toward the direction of the cluster.

The ratio of total to selective extinction ($R_{V}$) reflects the mean size of grains that are responsible for extinction as well as polarization. The value of $R_V$ ranges from $\approx$ 3.0 for diffuse ISM to $\approx$ 5.5 for dark clouds. Earlier studies were consistent with the assumption that the growth of $R_{V}$ is accompanied by the corresponding growth of $\lambda_{max}$  \citep{1978A&A....66...57W}. The correlation between the value of $R_{V}$ and $\lambda_{max}$ was studied by \citet{1978A&A....66...57W}, which is given as $R_{V} = (5.6\pm0.3) \times \lambda_{max}$. 
Using our derived value of $\lambda_{max}$, we have estimated $R_V$ as 3.02$\pm$0.20 for  the cluster NGC 1817. This suggests that the grain size in the line of sight of the cluster NGC 1817  is similar to that of the ISM.

\subsection{Polarization Efficiency} \label{eff}
\begin{figure*}
	\centering
\subfigure[The variation of maximum polarization with the color excess.]{\includegraphics[width=0.49\columnwidth]{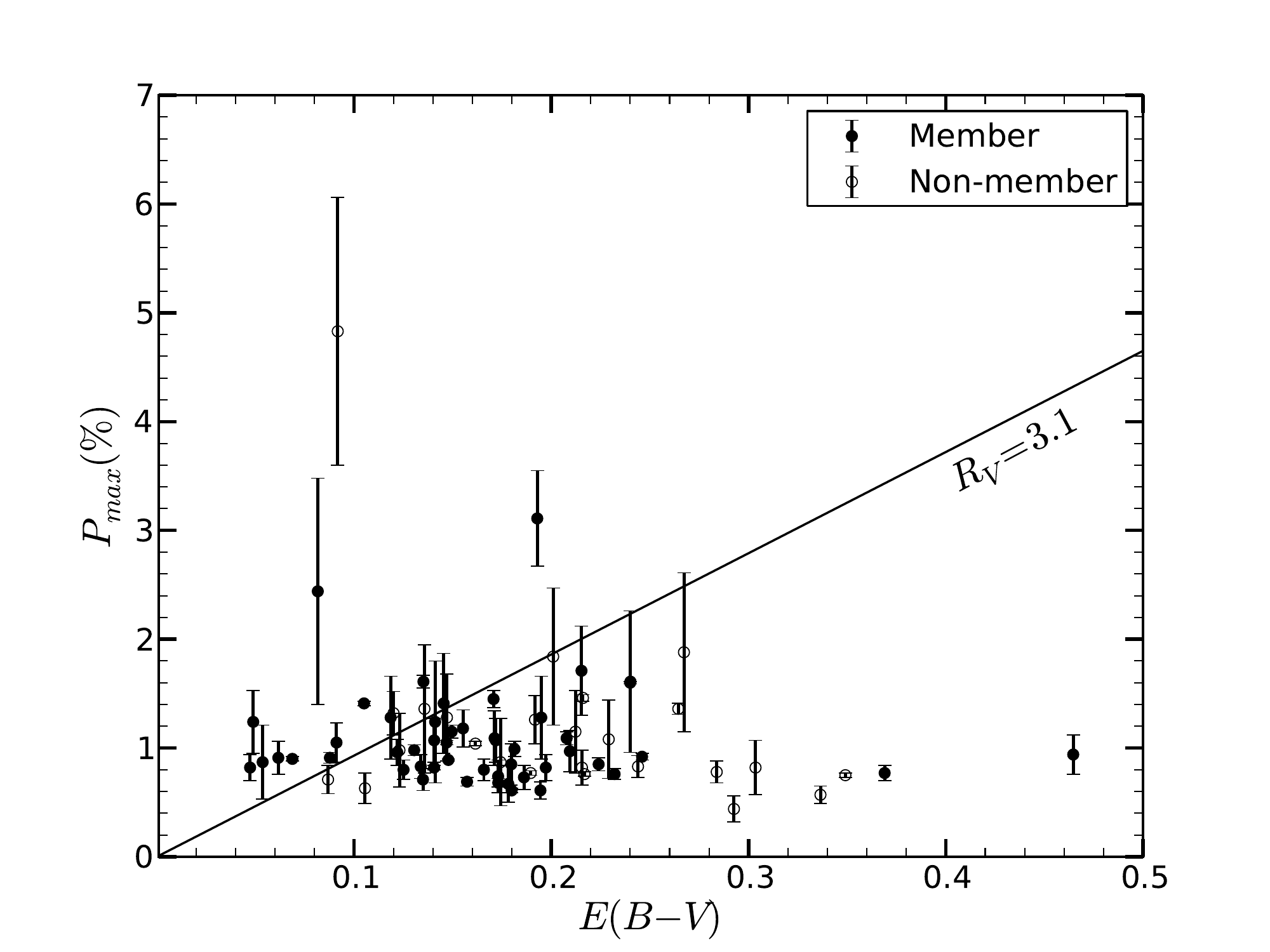}\label{fig:eff1}}  
\subfigure[The polarization efficiency as a function of color excess.]{\includegraphics[width=0.49\columnwidth]{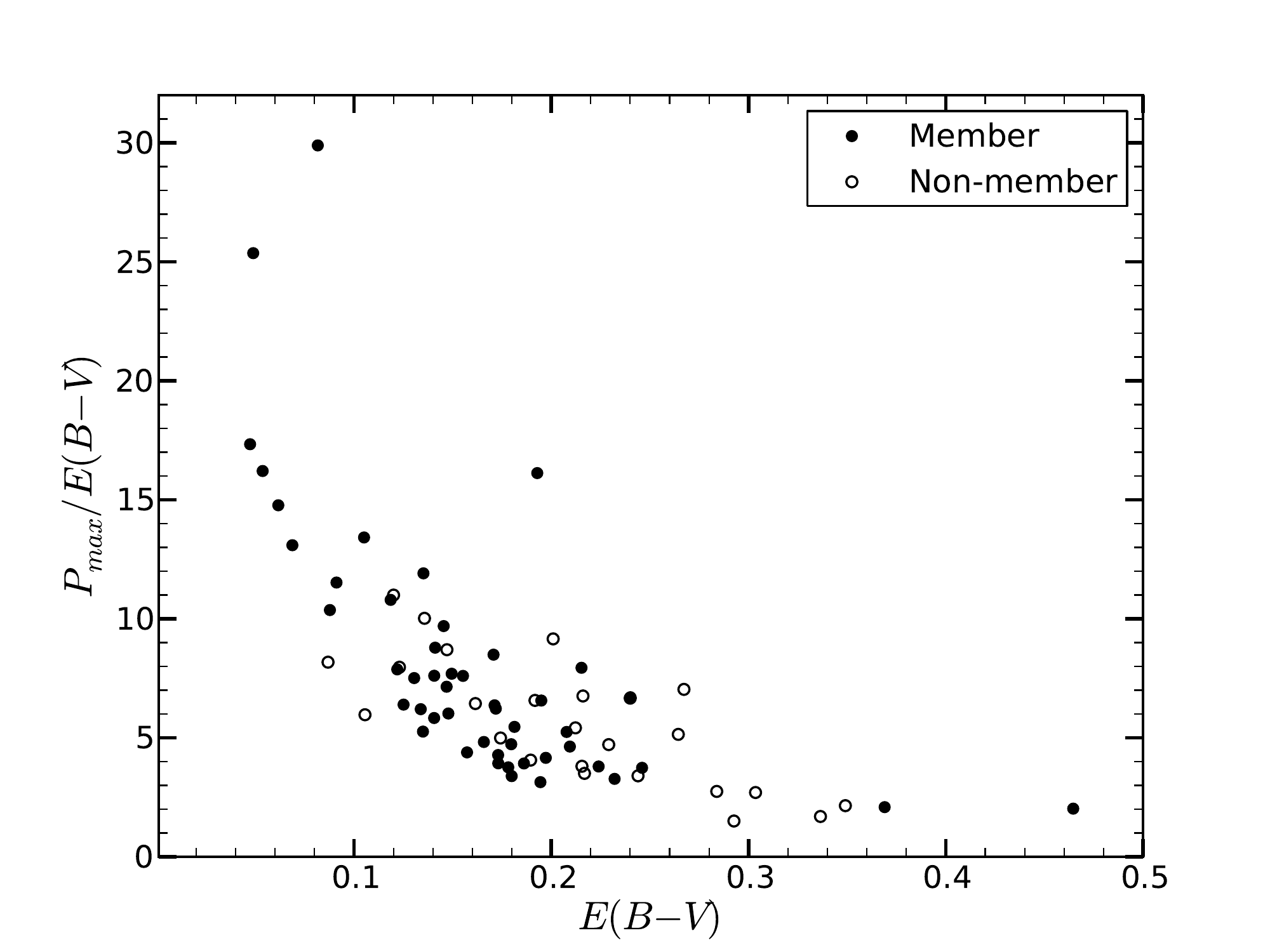}\label{fig:eff2} } 
    \caption{The polarization efficiency diagrams. The filled circles and open circles are for members and nonmembers, respectively. In panel (a), the straight line represents the line of maximum efficiency for $R_{V}$ = 3.1.}
    \label{fig:eff}
\end{figure*}

The polarizing efficiency of dust grains depends on the alignment, magnetic field strength, and its orientation \citep{2008JQSRT.109.1527V}. It is measured by the maximum polarization produced from a given amount of extinction.  The upper limit for maximum polarization is given by  \citet{1956ApJS....2..389H} as $P_{max}\leq 3.0 R_{V} \times E(B-V)$ and was found to be  2.5 \% using the mean value of $E(B-V)$ of 0.27 mag  \citep{2011AJ....142...59J} for the cluster NGC 1817. The average value of $P_{max}$ as derived in section \ref{ser} was found to be less than this value.  For the majority of the stars, the value of $P_{max}$ ( see Table \ref{tab:serkowski}) was found to be less than the maximum expected value of  $P_{max}$ in the direction of the cluster NGC 1817. For stars 18 and 54, the $P_{max}$ was found more than the expected maximum value.  Figure \ref{fig:eff} shows the polarization efficiency diagram. In Figure \ref{fig:eff1}, the straight line represents the line of maximum polarization efficiency corresponding to $R_{V}$ = 3.1. The majority of the stars lie below or within the maximum efficiency line whereas stars 18, 54, and 65 are located above the maximum efficiency line. A few stars with $E(B-V)$ $<$ 0.1 were also found slightly above the maximum efficiency line probably due to the large uncertainty in the $E(B-V)$ estimation.  We have not found any suitable reason for these stars, so we assume that it may be due to the excess affected by error or excess may be underestimated as the reddening we have taken from \citet[][]{2019A&A...628A..94A} with the 50th percentile value. Figure \ref{fig:eff2} is a plot  between polarization efficiency ($P_{max}/E(B-V)$) and  the reddening. Star 18 is not shown in this plot due to its high value of polarization efficiency. A clear decreasing trend of polarization efficiency was observed with an increased value of reddening.  For the majority of the stars, the value of $P_{max}$ was found to be almost constant.
  
\citet{2002ApJ...564..762F} have given a new estimation of the average efficiency as $P_{V}(\%) = 3.5 E(B-V)^{0.8}$ for $E(B-V)$ $<$ 1.0 mag. According to this relation, we found $P_{V}$ as 1.23 \% for the average value of $E(B-V)$ = 0.27 mag, which is very similar to the observed  $P_{V}$ or $P_{max}$. This also shows that  the polarizing efficiency of dust grains toward the cluster NGC 1817 is comparable with the average value observed in our Galaxy.

\section{Discussion} \label{sec:diss}
We have carried out multiband linear  polarimetric observations of 125  stars toward  the cluster NGC 1817 for the first time. Polarization is found to be wavelength-dependent for the majority of the stars.  Both  $P$ and $\theta$  show similar distributions for  members and nonmembers, which implies that the source of polarization for both types of stars is the same. The  $\theta$  for the majority of the stars in  the field of NGC 1817 are found to be not parallel to the  GP.  In general the polarization vectors are found to be nearly parallel to the GP \citep[e.g.][]{2004A&A...419..965M,2007MNRAS.378..881M,2010MNRAS.403.1577M,2007A&A...462..621V,2010A&A...513A..75O,2011MNRAS.411.1418E,2020AJ....159...99S}. However,  in some studies two or more types of polarization vectors  are found in which one is parallel to the GP  and the others are showing a different orientation than  the GP \citep{2008MNRAS.391..447F,2008MNRAS.388..105M}.  Also, there are some other  works where the polarization vectors are found  not to be  aligned with the GP \citep{2010MNRAS.403.2041V,2018RMxAA..54..293V}.  This nonalignment of the  polarization vector with the GP can give  a hint about the dust that  has been perturbed at the place \citep[e.g.][]{1978Ap&SS..54..425E}. The dust has been recently perturbed so particles do not have enough time to relax and orient further in the direction of the GP.  However, dust particles in most aged clusters are expected to have enough time to relax and polarize the light in the same orientation as the GP in the region \citep{1976MNRAS.177..499A}. The presence of local clouds in the vicinity of the cluster may be also responsible for the different orientation of the polarization vector than the GP.   We have noticed that 9 clouds  are located within a 5\degr\, radius from the center of the cluster NGC 1817 \citep[see][for details]{2005PASJ...57S...1D}. Cloud no. 4586  ($l$=189\degr 26\arcmin, $b$ = -10\degr 20\arcmin) was located nearest to the cluster NGC 1817 within 3\degr.5 radius from the cluster center. This cloud  is extended in 26.56 arcmin$^{2}$ region, whereas the cloud no. 4534 ($l$=182\degr 09\arcmin, $b$= -17\degr  57\arcmin) is dense and extended in a 3.6 degree$^{2}$ field of view \citep[see][]{2005PASJ...57S...1D,2011PASJ...63S...1D}.  \citet{2002ApJ...564..762F} explained this nonparallelism as the effect of a large random component of the magnetic field. 

In addition to this we have noticed that the position angles were more dispersed after $\sim$ 5\arcmin\, along the radial distance from the cluster center which is close to the core radius of the cluster.  The highly scattered polarization and position angle in the coronal region indicate that the  dust distribution is more diverse than that  for the core region of the cluster.
It appears that the dust particles are of different generations with diverse  components of the magnetic field, resulting in the highly scattered $\theta_{V}$  in the radial plot.  
Up to a cluster radius of 5\arcmin\,, the distribution of  $\theta_{V}$ is almost constant indicating that the magnetic field, which is responsible for the alignment of grains in the core region, is significantly stronger than that in the coronal region. 

If any dust layer(s) is present  at any distance in front of the cluster, the degree of polarization and extinction  show an increase in their values at that particular distance when plotted against the distance. Thus the number of sudden enhancement in degrees of polarization and extinction with distance are associated with the number of dust layers in the line of sight. Such distributions of the degree of polarization and hence the presence of dust layer(s) is found for many clusters \citep[e.g. in][]{1999AJ....117.2882W,2004A&A...419..965M,2008MNRAS.391..447F,2008MNRAS.388..105M,2010MNRAS.403.2041V,2018RMxAA..54..293V,2011MNRAS.411.1418E,2012MNRAS.419.2587E}. In the case of cluster NGC 1817, there is no indication for the presence of any major dust layer as the polarization as well as extinction have not shown any sudden increase in the distribution with the  distance. 

The average value of $\lambda_{max}$ derived  for the cluster NGC 1817 is found to be similar to that for the general ISM  \citep{1975ApJ...196..261S}. This shows that the size of the dust grains toward the line of sight of the cluster NGC 1817 is similar to that of the general ISM. The average values of $P_{max}$ and $\lambda_{max}$ are also found to be similar for  members and  nonmembers, respectively. The same value of $\lambda_{max}$ for members and nonmembers also indicates that light from both members and nonmembers are coming across the same population of foreground dust grains. It was found that the polarizing efficiency toward the cluster NGC 1817 is due to ISM dust grains. The decreasing trend of polarizing efficiency with the increase in color excess is also noticed, which may be because of the increase in the size of dust grain or a small change in the polarization position angle with an increase in reddening. 

Based on the Stokes plane (see section \ref{qu} and Figure \ref{fig:QU}) two sets of stars with different polarimetric characteristics appear to be present in the region of the cluster NGC 1817. Stars from these two classes are randomly located all over the  cluster region. We could not  discriminate these groups of  stars based on their distance and spatial distribution. In group 1 the degree of polarization and position angle are observed to have a little higher values. However, no separate distribution of $P$ and $\theta$  is seen for these two groups in Figures \ref{fig:pv_hist} and \ref{fig:thv_hist}. The position angle for group 1 is more aligned to the GP than that from group 2. It can be the case that there is a presence of a nonrelaxed cloud lying closer to the Sun that covers the cluster, which is far away and  consists of some nonmember stars, that may be responsible for the dominant component of the polarization for group 2.  Whereas the different polarization characteristics of group 1 could be due to the superposition of polarization due to this cloud and other component.  In order to get the polarization property of the additional component, we have subtracted the  Stokes parameters of group 2 from the respective Stokes parameters of group 1. As a result we obtained the values of $P$ and $\theta$ for the additional component as 0.38 \% and 145\degr, respectively. This derived value of $\theta$ is found to be similar to the angle for the GP. This  clearly indicates that the behind the first closest cloud  most of the layers of the ISM are relaxed with an orientation of the GP and a combination of these two is responsible for the polarization in group 1.  Also the first closest cloud has less dense parts which will  generate  a different composition of the polarimetric vectors for different stars  and show the continuous distribution of $\theta$ in Figure \ref{fig:thv_hist}.

In the present study some stars are identified with different polarization parameters. A few  nonmember stars (IDs: 18, 35, 44, 49, and 62) are found with a higher value of $P_{V}$. Star 18 is located at the cluster distance and shows the indication of intrinsic polarization  as we obtained a very small value of $\lambda_{max}$. Star 44 is located far away and behind the cluster explaining its higher polarization. The other three stars are foreground stars in the cluster and may have an intrinsic component of polarization resulting in a higher value of $P_V$.   Another two background  stars,  53 and 58, are observed with  a high value for the position angle (Figure \ref{fig:pv_thev}) and a low value of $\lambda_{max}$  (see Table \ref{tab:serkowski}).  This could also be due to  the  intrinsic polarization in it (see section \ref{ser}). The low degree of polarization in star 58 could be due to the  depolarization  of the intrinsic and ISM-induced  polarization vectors. 
The member star 32 also shows a similar behavior to these two nonmember stars. Some other stars (IDs: 52, 110, and 112) have shown relatively lower position angles than those of other stars. Nonmember stars 52 and 110 also show a variation of $\theta$ among the passbands and fitting of the Serkowski law resulted in large errors in the derived parameters. The star 114 has  very low values of $P_{B}$ and $P_{V}$. This star is at the cluster distance, also $\theta_{V}$ is near the GP. Again depolarization could be a reason for the low value of the $P_{V}$ in this star. However, with only two band observations we cannot say anything conclusive about the intrinsic nature of polarization in the star 114 and also for 112.

We have found 11 such members which show an intrinsic component in their polarization and six (IDs: 9, 13, 24, 27, 54, and 97) of them are giant stars. The giant stars are the candidates to have a noninterstellar component in their polarization, as these are evolved stars and hence have an extended atmosphere. The SB star 27 has shown a lower value of $\theta$ and also shows a dispersion in $\theta$ in different observed bands. The variation of the  polarization position angle with wavelength could be  due to the binary nature of this star. The higher $\sigma_{1}$ is also indicative of the presence of intrinsic polarization. The light from the binary star also gets polarized by the reflection from the atmosphere of the companion star. Also this can lead to variability of the polarization signal with time. A time-series polarimetric observation may shed light on the observed nature of the polarization in these stars. We have also calculated the average values of $P_{V}$ and $\theta_{V}$ for different types of variable and other  stars in the cluster NGC 1817 and we did not found any significant difference among them. These values are given in Table \ref{tab:var_avg}. The $P_{V}$ for pulsating variables  is found to be slightly smaller than that for  nonvariable  member stars. Also the average value of  $\theta_V$ for blue stragglers is found to be more than for other stars of the cluster.  This indicates that the depolarization had happened in these variables due to the superposition of their intrinsic polarization with the ISM polarization.  

\begin{figure*}
	\centering
\includegraphics[trim=5cm 0cm 5cm 0cm,width=\columnwidth]{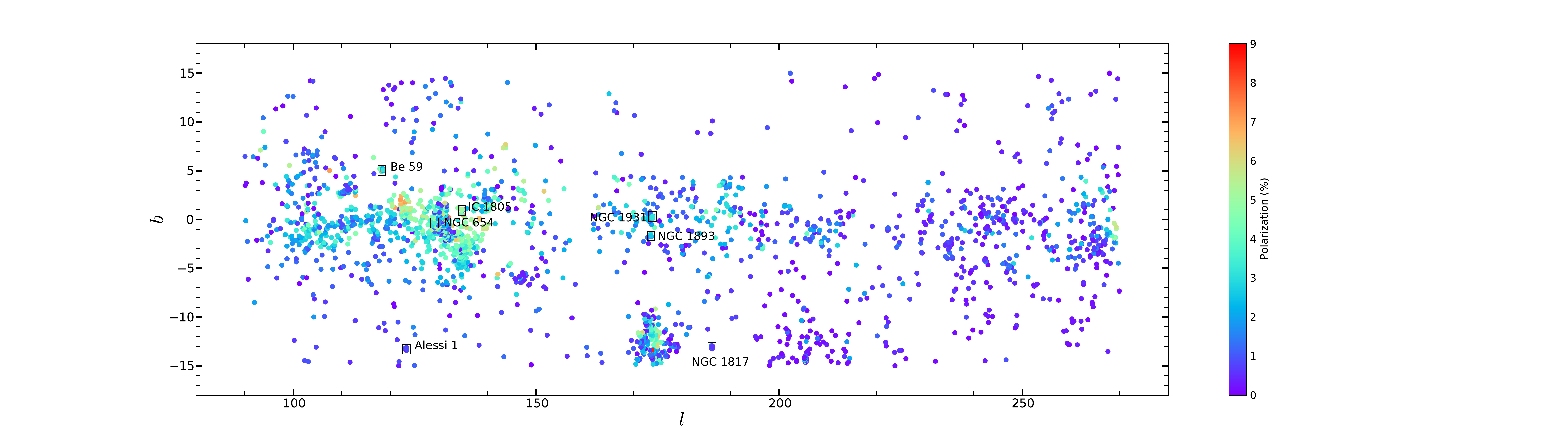}
\caption{The distribution of the degree of polarization in Galactic longitude (from 90\degr\, to 270\degr) and Galactic latitude (-15\degr\, to +15\degr). The data is extracted from \citet{2000AJ....119..923H}. Cluster NGC 1817 stars and clusters mentioned in section \ref{sec:intro} are also overplotted on this.}
    \label{fig:heiles}
\end{figure*}

In order to see the polarization behavior toward the Galactic anticenter direction (90\degr $<$ $l$ $<$ 270\degr; -15\degr $<$ $b$ $<$ 15\degr), the distribution of the degree of  polarization (taken from \citet{2000AJ....119..923H}) in Galactic longitude and latitude is shown in Figure \ref{fig:heiles} with the color bar for the degree of polarization on the right. We have selected only those stars for which the degree of polarization was above the 2$\sigma$ level.  The degrees of polarization for the stars of the cluster NGC 1817 are overplotted along with other observed clusters. The positions for these clusters  are marked by boxes along  with their name. There are a large number of observations near the Galactic plane but near to the position of cluster NGC 1817 a very few observations were carried. A relatively larger value of the degree of polarization  is found  near the galactic disk, which decreases while  going away from the galactic plane. There are indications for the presence of dust layers in front of the clusters (e.g. Be 59, NGC 1893, and NGC 654) that are near to the galactic disk, whereas for the clusters located away from the galactic disk no major dust layers are found (e.g. Casado Alessi 1 and NGC 1817).  For the clusters IC 1805 and NGC 1931 the information  about the dust layer was not available.  For the cluster NGC 1931 we have used polarization data from \citet{2013ApJ...764..172P} and the distances of the stars are taken from the Gaia DR2 data  in order to see the distribution of polarization with distance. We identified two dust layers near the distance of 800 pc and 2.6 kpc in the direction of the cluster NGC 1931. However, for the cluster IC 1805, we could not extract this information due to the unavailability of the exact position of polarimetrically observed stars in the cluster.

\section{Summary and Conclusions}\label{sec:sum}
The linear polarimetric study of 125 stars in the region of the cluster NGC 1817 is carried out in the B, V, R, and I bands. The average value of polarization and position angle are found to be 0.75\% $\pm$ 0.02\% and 133\degr.5$\pm$0\degr.8; 0.95\% $\pm$ 0.02\% and 119\degr.2$\pm$0\degr.5; 0.88\% $\pm$ 0.02\% and 117\degr.5$\pm$0\degr.6; and 0.51\% $\pm$ 0.01\% and 119\degr.8$\pm$0\degr.8 in bands B, V, R, and I, respectively. 
The source of the polarization is found to be same for member and nonmember stars as the distribution of $P$ and $\theta$ is almost similar in all of the bands. The members have an average value of $P_{V}$ and $\theta_{V}$ of 0.91\% and 118\degr.6 whereas  for nonmembers these values are 1.01\% and 120\degr.0.  
Highly scattered values of $P_{V}$ and $\theta_{V}$ are noticed in the coronal region of the cluster  specifying the diverse distribution of dust. The deviation of the polarization position angle from the GP is found and indicates that the dust has been disturbed at the place. We have identified two different groups of stars in the cluster which have different polarimetric characteristics. The first and second group correspond to $P_{V}$ and $\theta_{V}$ as 1.01\% and 124\degr.8 and 0.76\% and 115\degr.3, respectively.  It appears that both nonrelaxed cloud  and relaxed ISM layers are responsible for the polarization of group 1 stars whereas only the nonrelaxed cloud is responsible for the polarization of stars in group 2. The polarization of the majority of stars is found to be  ISM originated. The average values of  $P_{max}$  for members is found to be 0.93$\pm$0.03 \%. Also  the polarizing efficiency of dust grains is found to be  close to the average value due to the general ISM. The average value of $\lambda_{max}$ of 0.54$\pm$0.02 $\mu m$ reveals that the size distribution of dust grains toward the cluster NGC 1817 is similar to that of the general ISM. There are a few stars which may have intrinsic  component of  polarization. The lower value of the degree of polarization indicates that the starlight has been depolarized  due to the superposition of more than one component of polarization. The indication for the presence of dust layers were noticed in front of those clusters which lie near to galactic plane.

\section*{Acknowledgment}
We thank the referee for reading our paper and for the useful suggestions. We thank Dr. Biman J. Medhi for his support. Part of this work has made use of data from the European Space Agency (ESA) mission Gaia (https://www.cosmos.esa.int/gaia), processed by the Gaia Data Processing and Analysis Consortium (DPAC, https://www.cosmos.esa.int/web/gaia/dpac/consortium). Funding for the DPAC has been provided by national institutions, in particular, the institutions participating in the Gaia Multilateral Agreement.

\bibliography{ref}{}
\bibliographystyle{aasjournal}

\begin{longrotatetable}
\begin{deluxetable*}{llccccccccr}
\centering
\tabletypesize{\scriptsize}
\tablecaption{Results of the Broadband Polarimetric Observations of All Stars toward the Cluster NGC 1817.}
\label{tab:pthe}
\setlength\tabcolsep{4.0pt}
\tablehead{
\colhead{ID} & \colhead{Gaia ID} & \colhead{Offset(\arcsec)} &  \colhead{$P_{B}\left(\%\right)$} & \colhead{$\theta_{B}\left({^o}\right)$} & \colhead{$P_{V}\left(\%\right)$} & \colhead{$\theta_{V}\left({^o}\right)$} & \colhead{$P_{R}\left(\%\right)$} & \colhead{$\theta_{R}\left({^o}\right)$} & \colhead{$P_{I}\left(\%\right)$} & \colhead{$\theta_{I}\left({^o}\right)$}
}

\startdata
1*$^{a}$     &   3394757273339924864 &  0.16  &     0.94$\pm$0.04  &  139.2$\pm$ 1.4  &  1.08$\pm$0.03  &  128.3$\pm$ 0.7  &  1.01$\pm$0.01  &  126.3$\pm$ 0.1  &  0.78$\pm$0.13  &  125.8$\pm$ 4.8  \\
2*     &   3394757028525131776 &  0.02  &     0.82$\pm$0.01  &  155.9$\pm$ 0.2  &  0.99$\pm$0.10  &  129.2$\pm$ 2.8  &  1.00$\pm$0.03  &  121.0$\pm$ 0.9  &  0.58$\pm$0.13  &  128.5$\pm$ 6.3  \\
3     &   3394757135900949504 &  0.47  &     0.35$\pm$0.12  &  178.0$\pm$10.0  &  0.75$\pm$0.04  &  120.6$\pm$ 1.6  &  1.01$\pm$0.04  &  116.8$\pm$ 1.2  &  0.46$\pm$0.12  &  141.1$\pm$ 7.2  \\
4*     &   3394756276907502080 &  0.33  &     0.80$\pm$0.17  &  141.6$\pm$ 6.1  &  0.95$\pm$0.05  &  118.2$\pm$ 1.4  &  0.65$\pm$0.01  &  117.6$\pm$ 0.4  &  0.36$\pm$0.14  &  132.2$\pm$11.4  \\
5*     &   3394756173828291968 &  0.28  &     0.85$\pm$0.04  &  159.4$\pm$ 1.3  &  0.50$\pm$0.08  &  112.6$\pm$ 4.8  &  0.63$\pm$0.02  &  109.2$\pm$ 0.8  &  0.47$\pm$0.13  &  112.8$\pm$ 7.6  \\
6     &   3394756620504902912 &  0.16  &             -      &  	 -	       &  1.09$\pm$0.11  &  125.2$\pm$ 2.8  &  1.05$\pm$0.21  &  114.6$\pm$ 5.7  &  0.56$\pm$0.24  &  117.2$\pm$12.2  \\
7     &   3394755619782056064 &  0.41  &     0.69$\pm$0.01  &  149.4$\pm$ 0.1  &  0.76$\pm$0.01  &  132.0$\pm$ 0.1  &  0.66$\pm$0.12  &  120.8$\pm$ 5.3  &  0.46$\pm$0.06  &  128.7$\pm$ 3.7  \\
8*     &   3394755349194605440 &  0.51  &     0.75$\pm$0.06  &  176.5$\pm$ 2.2  &  0.64$\pm$0.10  &  119.8$\pm$ 4.5  &  	-	      &  	   -          &      -      	&      -  \\
9*$^{a}$     &   3394737722648827136 &  0.54  &     0.93$\pm$0.06  &  143.9$\pm$ 1.9  &  0.94$\pm$0.06  &  128.9$\pm$ 1.8  &  0.68$\pm$0.23  &  122.4$\pm$ 9.7  &  0.19$\pm$0.07  &  167.0$\pm$10.8  \\
10*$^{a}$    &   3394743563804338944 &  0.55  &     0.59$\pm$0.15  &  160.2$\pm$ 7.3  &  0.82$\pm$0.09  &  114.6$\pm$ 3.0  &  0.57$\pm$0.13  &  113.5$\pm$ 6.7  &  0.33$\pm$0.06  &  113.7$\pm$ 5.1  \\
11*$^{b}$   &   3394743563804338048 &  0.43  &     0.60$\pm$0.18  &  137.7$\pm$ 8.8  &  0.76$\pm$0.18  &  120.4$\pm$ 6.8  &  0.66$\pm$0.12  &  104.1$\pm$ 5.3  &      -      	&       -	   \\
12*$^{a}$    &   3394755589712747648 &  0.37  &     0.79$\pm$0.05  &  161.2$\pm$ 1.9  &  0.89$\pm$0.01  &  124.3$\pm$ 0.3  &  0.86$\pm$0.06  &  118.1$\pm$ 1.9  &  0.52$\pm$0.13  &  124.0$\pm$ 7.3  \\
13*$^{a}$    &   3394755520993275904 &  0.52  &     0.54$\pm$0.01  &  128.3$\pm$ 0.1  &  0.89$\pm$0.01  &  125.6$\pm$ 0.2  &  0.75$\pm$0.14  &  116.3$\pm$ 5.2  &  0.53$\pm$0.01  &  117.0$\pm$ 0.7  \\
14*$^{b}$    &   3394743731306390016 &  0.46  &    	-           &  	 -	       &  0.53$\pm$0.13  &  132.4$\pm$ 6.8  &  0.97$\pm$0.07  &  119.4$\pm$ 2.1  &  0.36$\pm$0.06  &  105.7$\pm$ 4.4  \\
15*    &   3394755486633539584 &  0.42  &     	-           &  	-	       &  0.77$\pm$0.06  &  122.3$\pm$ 2.3  &  0.74$\pm$0.01  &  111.4$\pm$ 0.3  &  0.45$\pm$0.15  &  116.9$\pm$ 9.5  \\
16*    &   3394755791574866432 &  0.26  &    	-           &  	-	       &  0.85$\pm$0.05  &  107.8$\pm$ 1.7  &  0.92$\pm$0.33  &   99.4$\pm$10.4  &  0.49$\pm$0.08  &  145.7$\pm$ 4.6  \\
17*    &   3394755589712750720 &  0.35  &     1.42$\pm$0.07  &  148.2$\pm$ 1.5  &  1.42$\pm$0.37  &  111.5$\pm$ 7.5  &  1.22$\pm$0.25  &  123.5$\pm$ 5.8  &  0.69$\pm$0.18  &  130.7$\pm$ 7.5  \\
18    &   3394755864590660096 &  0.17  &     2.33$\pm$0.77  &  122.5$\pm$ 9.8  &  1.81$\pm$0.71  &  109.8$\pm$11.2  &  0.97$\pm$0.06  &  110.3$\pm$ 1.9  &  0.52$\pm$0.22  &  160.1$\pm$12.2  \\
19*$^{c}$    &   3394744177982986752 &  0.45  &     0.89$\pm$0.07  &  140.6$\pm$ 2.1  &  1.13$\pm$0.03  &  131.8$\pm$ 0.7  &  1.01$\pm$0.13  &  116.8$\pm$ 3.8  &  0.76$\pm$0.10  &  126.1$\pm$ 3.8  \\
20    &   3394756414346436736 &  0.43  &     	-           &  	 -	       &  0.91$\pm$0.06  &  116.3$\pm$ 2.0  &  0.53$\pm$0.11  &  117.0$\pm$ 6.0  &  0.54$\pm$0.22  &  127.5$\pm$11.4  \\
21    &   3394756139468531712 &  0.59  &     	-           &  	-	       &  	   -          &      -      &  1.37$\pm$0.11  &  107.2$\pm$ 2.4  &  0.41$\pm$0.09  &   93.8$\pm$ 6.3  \\
22$^{b}$    &   3394743873041962752 &  0.22  &     	-           &  	-	       &  	  -           &       -     &  0.90$\pm$0.17  &  123.4$\pm$ 5.5  &  0.39$\pm$0.17  &  132.7$\pm$12.4  \\
23*$^{b}$    &   3394744216639326208 &  0.13  &     0.67$\pm$0.07  &  153.1$\pm$ 2.8  &  0.83$\pm$0.05  &  117.7$\pm$ 1.8  &  0.83$\pm$0.17  &  118.7$\pm$ 5.8  &  0.54$\pm$0.06  &  125.8$\pm$ 3.0  \\
24*$^{a}$    &   3394743838682218368 &  0.30  &     0.71$\pm$0.07  &  145.8$\pm$ 2.9  &  0.91$\pm$0.01  &  126.7$\pm$ 0.2  &  0.89$\pm$0.20  &  108.0$\pm$ 6.4  &  0.49$\pm$0.01  &  113.3$\pm$ 0.3  \\
25*$^{b}$    &   3394743804322483456 &  0.43  &    	-           &  	-	       &  0.92$\pm$0.13  &  112.2$\pm$ 4.1  &  0.89$\pm$0.01  &  108.3$\pm$ 0.2  &  0.88$\pm$0.15  &  103.6$\pm$ 4.8  \\
26*$^{b,d}$    &   3394757651297008640 &  0.38  &     0.56$\pm$0.14  &   98.8$\pm$ 7.3  &  0.74$\pm$0.09  &  102.8$\pm$ 3.6  &  0.66$\pm$0.15  &   80.6$\pm$ 6.3  &  0.57$\pm$0.01  &  103.4$\pm$ 0.6  \\
27*$^{a}$    &   3394745797187275392 &  0.18  &     0.70$\pm$0.03  &  126.4$\pm$ 1.3  &  1.22$\pm$0.01  &   90.1$\pm$ 0.1  &  0.85$\pm$0.10  &  104.1$\pm$ 3.4  &  0.52$\pm$0.08  &  100.0$\pm$ 4.6  \\
28    &   3394745724171212160 &  0.02  &     0.65$\pm$0.16  &   83.3$\pm$ 6.9  &  0.78$\pm$0.05  &  140.0$\pm$ 1.8  &  0.69$\pm$0.09  &  137.3$\pm$ 3.9  &  0.30$\pm$0.15  &  114.1$\pm$14.5  \\
29*$^{b}$  &   3394745797187279360   &  0.20   &     	-    &   	 -     &  0.98$\pm$0.19  &    99.6$\pm$5.6  &  0.94$\pm$0.11  &  106.4$\pm$3.3  &  0.66$\pm$0.20  &  121.8$\pm$8.6  \\
30    &   3394762255501947776 &  0.39  &   	-           &  	-	       &  0.97$\pm$0.28  &  105.1$\pm$ 8.5  &  1.08$\pm$0.09  &  120.4$\pm$ 2.6  &	     -     	&    -  	   \\
31$^{e}$    &   3394762186782453376 &  0.67  &     0.73$\pm$0.18  &   71.9$\pm$ 6.9  &  0.78$\pm$0.31  &  106.8$\pm$11.6  &         -	      &  	  -	      &      -    	&    -  	   \\
32*    &   3394757479498332928 &  0.37  &     1.09$\pm$0.34  &  153.4$\pm$ 8.8  &  0.86$\pm$0.38  &  155.2$\pm$12.7  &  0.70$\pm$0.11  &  120.0$\pm$ 4.7  &     -	        &    -  	   \\
33    &   3394757303403446144 &  0.64  &     1.83$\pm$0.73  &  113.2$\pm$11.8  &  0.98$\pm$0.05  &   94.4$\pm$ 1.5  &  1.01$\pm$0.22  &  145.4$\pm$ 6.5  &     -	        &    -  	   \\
34*$^{d}$    &   3394757509561465344 &  0.20  &     0.35$\pm$0.01  &   91.4$\pm$ 0.1  &  1.18$\pm$0.02  &  126.8$\pm$ 0.5  &  0.98$\pm$0.17  &  126.9$\pm$ 5.1  &     -	        &    -  	   \\
35    &   3394757410778858368 &  0.24  &     1.59$\pm$0.15  &  115.8$\pm$ 2.7  &  1.60$\pm$0.59  &  160.4$\pm$10.4  &  1.39$\pm$0.49  &  113.5$\pm$10.4  &     -	        &    -  	   \\
36*    &   3394759025686541568 &  0.08  &     	-           &  	-	       &  1.14$\pm$0.22  &  118.1$\pm$ 5.7  &  0.68$\pm$0.30  &  111.3$\pm$13.1  &     -	        &    -  	   \\
37    &   3394759025686542336 &  0.19  &    	-           &  	-	       &  0.94$\pm$0.24  &  120.8$\pm$ 7.4  &  0.61$\pm$0.12  &  116.7$\pm$ 5.9  &     -	        &    -  	   \\
38    &   3394762083703243904 &  0.15  &     0.72$\pm$0.15  &  125.9$\pm$ 6.2  &  0.78$\pm$0.07  &  115.1$\pm$ 2.5  &  0.67$\pm$0.18  &  111.7$\pm$ 7.8  &     -	        &    -  	   \\
39    &   3394759025686536704 &  0.54  &     	-           &  	 -	       &  1.36$\pm$0.01  &  111.6$\pm$ 0.2  &  1.30$\pm$0.22  &  115.5$\pm$ 5.1  &     -	        &    -  	   \\
40*$^{b}$    &   3394758922607329664 &  0.05  &     1.63$\pm$0.03  &  138.4$\pm$ 0.5  &  1.03$\pm$0.02  &  123.7$\pm$ 0.5  &  1.10$\pm$0.01  &  123.0$\pm$ 0.1  &     -	        &    -  	   \\
41*$^{c}$    &   3394758716448900224 &  0.21  &     0.39$\pm$0.13  &   88.0$\pm$ 9.5  &  0.81$\pm$0.01  &  120.6$\pm$ 0.2  &  0.67$\pm$0.04  &  128.1$\pm$ 1.7  &     -	        &    -  	   \\
42*$^{b}$    &   3394758716448904448 &  0.26  &     0.43$\pm$0.04  &  148.4$\pm$ 2.8  &  0.66$\pm$0.26  &  120.6$\pm$11.3  &  0.79$\pm$0.21  &  114.0$\pm$ 7.9  &     -	        &    -  	   \\
43    &   3394758682089159424 &  0.20  &     1.41$\pm$0.49  &   88.2$\pm$ 9.8  &  1.44$\pm$0.06  &  118.3$\pm$ 1.2  &  1.24$\pm$0.60  &  128.6$\pm$14.7  &     -	        &    -  	   \\
44    &   3394758647729422848 &  0.37  &     1.36$\pm$0.10  &   79.4$\pm$ 2.0  &  1.65$\pm$0.71  &   97.2$\pm$12.3  &  1.10$\pm$0.01  &  177.7$\pm$ 0.1  &     -           &    -  	   \\
45*$^{d}$    &   3394759094406004480 &  0.11  &     0.85$\pm$0.07  &  116.0$\pm$ 2.5  &  0.70$\pm$0.03  &  124.9$\pm$ 1.2  &  0.82$\pm$0.22  &  104.1$\pm$ 7.8  &  0.72$\pm$0.03  &  112.9$\pm$1.1	   \\
46*$^{b}$    &   3394758853887838720 &  0.08  &     0.80$\pm$0.06  &  132.2$\pm$ 2.2  &  0.72$\pm$0.01  &  117.6$\pm$ 0.4  &  0.57$\pm$0.08  &  129.0$\pm$ 4.4  &  0.46$\pm$0.23  &  133.1$\pm$14.5	   \\
47    &   3394758780871784704 &  0.34  &     0.92$\pm$0.07  &  148.1$\pm$ 2.1  &  1.08$\pm$0.04  &  137.1$\pm$ 1.1  &  0.64$\pm$0.06  &  112.9$\pm$ 2.6  &     -     	&   -		   \\
48*$^{a}$    &   3394759266204683904 &  0.21  &     0.55$\pm$0.10  &  103.2$\pm$ 5.5  &  0.61$\pm$0.03  &  129.5$\pm$ 1.5  &  0.62$\pm$0.10  &  124.2$\pm$ 4.6  &  0.41$\pm$0.15  &  112.9$\pm$10.3	   \\
49    &   3394759403643627904 &  0.16  &     0.77$\pm$0.15  &  157.8$\pm$ 5.6  &  1.74$\pm$0.08  &  101.7$\pm$ 1.3  &  1.53$\pm$0.13  &   96.0$\pm$ 2.4  &     -      	&   -		   \\ 
50    &   3394758819528096768 &  0.09  &     	-           &  	 -	       &  1.46$\pm$0.05  &  130.5$\pm$ 0.9  &  0.63$\pm$0.09  &   99.9$\pm$ 3.9  &     -     	&   -		   \\
51*    &   3394758063613853312 &  0.20  &     1.14$\pm$0.36  &  139.8$\pm$ 9.2  &  0.96$\pm$0.10  &  122.3$\pm$ 3.1  &  2.07$\pm$0.72  &  107.2$\pm$10.3  &  1.09$\pm$0.30  &   90.8$\pm$ 7.8  \\
52    &   3394754107947356160 &  0.57  &     0.64$\pm$0.01  &  177.7$\pm$ 0.1  &  1.07$\pm$0.43  &   90.1$\pm$11.5  &     -	      &   	-	      &  0.12$\pm$0.05  &  117.4$\pm$12.8  \\
53    &   3394753768646663168 &  0.93  &     1.01$\pm$0.39  &  149.8$\pm$11.2  &  1.08$\pm$0.02  &  166.7$\pm$ 0.6  &  0.91$\pm$0.38  &  126.0$\pm$12.0  &  0.56$\pm$0.19  &  157.0$\pm$ 9.8  \\
54*$^{a}$   &   3394750367032577152 &  0.28  &     	-           &  	 -	       &  0.92$\pm$0.03  &  118.3$\pm$ 1.0  &  0.57$\pm$0.09  &  112.9$\pm$ 4.6  &  0.31$\pm$0.08  &  109.9$\pm$ 7.4  \\
55    &   3394750332672841088 &  0.36  &     0.31$\pm$0.14  &  149.7$\pm$13.5  &  0.91$\pm$0.18  &  114.3$\pm$ 5.7  &  0.46$\pm$0.22  &  121.0$\pm$13.8  &     -      	&    -  	   \\
56    &   3394753837366131712 &  0.60  &     0.51$\pm$0.07  &  147.3$\pm$ 3.8  &  0.94$\pm$0.07  &  128.1$\pm$ 2.1  &  0.83$\pm$0.11  &  121.6$\pm$ 3.9  &  0.53$\pm$0.05  &  121.4$\pm$ 2.5  \\
57    &   3394760022119016960 &  0.64  &     0.87$\pm$0.25  &  117.3$\pm$ 8.1  &  	   -          &       -     &  	-	      &  	   -          &  0.65$\pm$0.16  &  144.4$\pm$ 7.2  \\
58    &   3394753837366127360 &  0.49  &     0.89$\pm$0.08  &  170.2$\pm$ 2.6  &  0.54$\pm$0.07  &  160.5$\pm$ 3.5  &  0.45$\pm$0.21  &  114.5$\pm$13.2  &     -     	&     -	         \\
59*    &   3394750710629942784 &  0.06  &     0.92$\pm$0.32  &  139.7$\pm$ 9.9  &  1.23$\pm$0.01  &  102.7$\pm$ 0.1  &  0.99$\pm$0.21  &  107.3$\pm$ 6.1  &  0.38$\pm$0.15  &  104.1$\pm$11.4  \\
60*    &   3394750504471515648 &  0.48  &     0.84$\pm$0.13  &  121.2$\pm$ 4.4  &  0.88$\pm$0.05  &  117.9$\pm$ 1.6  &  1.12$\pm$0.13  &  120.8$\pm$ 3.2  &  0.68$\pm$0.29  &  131.1$\pm$12.4  \\
61*$^{a}$    &   3394750603254043776 &  0.29  &     0.55$\pm$0.09  &  145.2$\pm$ 4.6  &  1.06$\pm$0.07  &  119.9$\pm$ 2.0  &  1.00$\pm$0.11  &  114.4$\pm$ 3.1  &  0.47$\pm$0.08  &  119.7$\pm$ 5.1  \\
62    &   3394750435752037888 &  0.88  &     1.22$\pm$0.40  &  162.4$\pm$ 9.3  &  1.67$\pm$0.42  &  117.6$\pm$ 7.2  &  0.44$\pm$0.21  &  118.5$\pm$13.3  &     -      	&     -  	   \\
63*    &   3394760159557957888 &  1.21  &     0.42$\pm$0.03  &  123.1$\pm$ 2.2  &  1.11$\pm$0.04  &  126.4$\pm$ 1.0  &  0.93$\pm$0.07  &  115.5$\pm$ 2.1  &  0.72$\pm$0.09  &  126.1$\pm$ 3.6  \\
64*    &   3394756929742552832 &  0.46  &     0.95$\pm$0.22  &  166.6$\pm$ 6.5  &  1.25$\pm$0.04  &  138.4$\pm$ 1.0  &  0.94$\pm$0.11  &  139.9$\pm$ 3.3  &  0.42$\pm$0.13  &  140.2$\pm$ 8.5  \\
65*    &   3394756547488780416 &  0.18  &     	-           &  	-	       &  1.33$\pm$0.02  &  123.2$\pm$ 0.5  &  0.85$\pm$0.05  &  130.9$\pm$ 1.9  &  0.80$\pm$0.10  &  118.1$\pm$ 3.6  \\
66*    &   3394756478769321472 &  0.53  &     1.09$\pm$0.21  &  129.9$\pm$ 5.6  &  1.11$\pm$0.42  &  120.5$\pm$10.9  &  0.88$\pm$0.21  &  116.6$\pm$ 6.9  &  0.41$\pm$0.17  &   80.7$\pm$11.5  \\
67    &   3394747309015893760 &  0.46  &     	 -          &  	-	       &  0.75$\pm$0.10  &  139.8$\pm$ 3.9  &  0.70$\pm$0.01  &  135.7$\pm$ 0.3  &  0.47$\pm$0.20  &  160.8$\pm$12.0  \\
68*    &   3394747343375624704 &  0.18  &     1.36$\pm$0.04  &  102.3$\pm$ 0.9  &  0.78$\pm$0.31  &  114.6$\pm$11.2  &  0.82$\pm$0.12  &  128.8$\pm$ 3.7  &  0.53$\pm$0.02  &  106.9$\pm$ 1.0  \\
69    &   3394735489265899136 &  0.18  &     0.43$\pm$0.06  &  142.3$\pm$ 4.2  &  0.88$\pm$0.11  &  128.0$\pm$ 3.7  &  0.97$\pm$0.07  &  125.7$\pm$ 2.1  &  0.52$\pm$0.10  &  116.6$\pm$ 5.5  \\
70    &   3394748820844371200 &  0.14  &     0.91$\pm$0.42  &  159.3$\pm$13.0  &  1.04$\pm$0.14  &  118.2$\pm$ 3.8  &  0.93$\pm$0.37  &  125.7$\pm$11.1  &  0.41$\pm$0.03  &  126.4$\pm$ 2.1  \\
71    &   3394735489265895424 &  0.23  &     	 -          &  	 -	       &  0.67$\pm$0.21  &  110.4$\pm$ 9.1  &  0.68$\pm$0.01  &  127.3$\pm$ 0.5  &      -     	&      - 	   \\
72    &   3394747270359425920 &  0.13  &     0.85$\pm$0.21  &  157.6$\pm$ 7.0  &  0.92$\pm$0.11  &  120.5$\pm$ 3.4  &  0.97$\pm$0.01  &  125.6$\pm$ 0.4  &  0.47$\pm$0.03  &  117.7$\pm$ 1.7  \\
73    &   3394735450609427584 &  0.12  &     0.45$\pm$0.21  &  122.1$\pm$14.0  &  0.97$\pm$0.06  &  118.4$\pm$ 1.7  &  1.23$\pm$0.12  &  123.7$\pm$ 2.7  &  0.41$\pm$0.04  &  119.4$\pm$ 3.0  \\
74    &   3394748747828175232 &  0.07  &     0.56$\pm$0.01  &  119.4$\pm$ 0.4  &  1.16$\pm$0.06  &  123.4$\pm$ 1.5  &  1.11$\pm$0.04  &  133.1$\pm$ 1.1  &  0.75$\pm$0.07  &  126.0$\pm$ 2.5  \\
75*    &   3394748752124891648 &  0.13  &     0.59$\pm$0.02  &  160.3$\pm$ 0.8  &  1.04$\pm$0.17  &  121.3$\pm$ 4.6  &  0.88$\pm$0.18  &  129.5$\pm$ 5.8  &  0.63$\pm$0.10  &  129.7$\pm$ 4.6  \\
76    &   3394749164441742720 &  0.49  &     0.42$\pm$0.09  &  127.4$\pm$ 6.6  &  0.78$\pm$0.12  &  112.5$\pm$ 4.0  &  0.83$\pm$0.02  &  120.7$\pm$ 0.6  &  0.50$\pm$0.10  &  114.9$\pm$ 5.7  \\
77    &   3394749263224249216 &  0.52  &     0.32$\pm$0.05  &  114.4$\pm$ 4.3  &  0.83$\pm$0.04  &  115.1$\pm$ 1.4  &  0.89$\pm$0.10  &  121.0$\pm$ 3.2  &  0.27$\pm$0.10  &  105.0$\pm$10.8  \\
78*    &   3394748958283307776 &  0.28  &     	-           &  	-	       &  0.61$\pm$0.27  &  100.3$\pm$12.8  &  0.84$\pm$0.12  &  100.8$\pm$ 3.6  &     -      	&      -  	   \\
79    &   3394737035454106880 &  0.21  &     0.75$\pm$0.02  &  179.9$\pm$ 0.8  &  0.68$\pm$0.03  &  119.0$\pm$ 1.4  &  0.63$\pm$0.11  &  124.8$\pm$ 5.0  &  0.40$\pm$0.06  &  111.7$\pm$ 4.7  \\
80    &   3394736966734628608 &  0.06  &     0.73$\pm$0.25  &  150.0$\pm$  9.5  &  1.02$\pm$0.17  &  121.9$\pm$ 4.7  &  0.89$\pm$0.03  &  119.5$\pm$ 1.1  &  0.55$\pm$0.04  &  101.8$\pm$ 1.8  \\
81*    &   3394737069813831808 &  0.07  &     0.46$\pm$0.04  &  154.8$\pm$ 2.3  &  0.69$\pm$0.09  &  109.7$\pm$ 3.9  &  0.83$\pm$0.05  &  123.5$\pm$ 1.7  &  0.50$\pm$0.21  &  119.0$\pm$12.2  \\
82*    &   3394749400663426304 &  0.15  &     	-           &  	-	       &  0.71$\pm$0.18  &  110.2$\pm$ 7.3  &  1.45$\pm$0.13  &  126.1$\pm$ 2.5  &  1.16$\pm$0.39  &  112.7$\pm$ 9.6  \\
83*    &   3394749439319617152 &  0.34  &     0.43$\pm$0.01  &  114.0$\pm$ 0.3  &  0.83$\pm$0.01  &  114.6$\pm$ 0.1  &  0.69$\pm$0.05  &  110.5$\pm$ 2.0  &  0.18$\pm$0.01  &  110.0$\pm$ 0.6  \\
84    &   3394737207252776704 &  0.30  &     	-           &  	-  	       &  1.06$\pm$0.01  &  121.7$\pm$ 0.1  &  0.76$\pm$0.25  &  128.0$\pm$ 9.4  &  0.24$\pm$0.05  &   67.0$\pm$ 6.1  \\
85*$^{b}$    &   3394743769962750592 &  0.48  &     0.49$\pm$0.16  &  136.0$\pm$ 9.2  &  0.98$\pm$0.05  &  115.9$\pm$ 1.3  &  0.90$\pm$0.01  &  117.3$\pm$ 0.5  &  0.41$\pm$0.09  &  115.5$\pm$ 6.4  \\
86*$^{b}$    &   3394756105108801664 &  0.42  &     0.45$\pm$0.17  &  139.6$\pm$10.8  &  0.92$\pm$0.01  &  120.8$\pm$ 0.2  &  0.86$\pm$0.02  &  117.3$\pm$ 0.5  &  0.59$\pm$0.13  &  126.9$\pm$ 6.1  \\
87*$^{b}$   &   3394756139468534400 &  0.43  &     0.70$\pm$0.30  &  114.4$\pm$12.3  &  0.91$\pm$0.06  &  129.9$\pm$ 2.0  &  0.82$\pm$0.04  &  110.3$\pm$ 1.3  &  0.56$\pm$0.25  &  178.9$\pm$12.6  \\
88*    &   3394744285358807040 &  0.51  &     0.95$\pm$0.20  &  178.9$\pm$ 6.0  &  1.15$\pm$0.27  &  115.6$\pm$ 6.8  &  0.88$\pm$0.01  &  107.6$\pm$ 0.3  &  0.47$\pm$0.02  &  112.5$\pm$ 1.2  \\
89*$^{b}$    &   3394744113560117504 &  0.44  &     1.24$\pm$0.34  &   99.6$\pm$ 7.9  &  1.36$\pm$0.13  &  112.8$\pm$ 2.8  &  1.42$\pm$0.21  &  122.2$\pm$ 4.2  &  1.34$\pm$0.41  &  139.9$\pm$ 8.8  \\
90*$^{c}$    &   3394742979688983680 &  0.24  &     0.55$\pm$0.21  &  108.1$\pm$10.8  &  0.75$\pm$0.02  &  119.8$\pm$ 0.8  &  0.78$\pm$0.12  &  113.9$\pm$ 4.3  &  0.47$\pm$0.06  &  117.2$\pm$ 3.4  \\
91$^{c}$    &   3394745827250425344 &  0.35  &     0.24$\pm$0.04  &  133.4$\pm$ 5.1  &  0.84$\pm$0.14  &  122.2$\pm$ 4.8  &  0.78$\pm$0.18  &  115.1$\pm$ 6.5  &  0.42$\pm$0.04  &  125.4$\pm$ 2.7  \\
92*$^{a}$    &   3394745522309372672 &  0.47  &     	-           &  	 -	       &  0.81$\pm$0.04  &  136.8$\pm$ 1.6  &  0.87$\pm$0.19  &  122.1$\pm$ 6.3  &  0.58$\pm$0.04  &  125.8$\pm$ 1.7  \\
93*$^{b}$   &   3394743976121154688 &  0.47  &     0.82$\pm$0.06  &  134.1$\pm$ 2.0  &  1.16$\pm$0.07  &  121.9$\pm$ 1.8  &  1.05$\pm$0.21  &  122.2$\pm$ 5.7  &  0.61$\pm$0.01  &  122.1$\pm$ 0.4  \\
94*$^{b}$    &   3394745522309371648 &  0.55  &     0.48$\pm$0.12  &   79.8$\pm$ 7.0  &  0.83$\pm$0.18  &  131.9$\pm$ 6.2  &  0.78$\pm$0.26  &  116.3$\pm$ 9.5  &  0.52$\pm$0.05  &  120.6$\pm$ 2.9  \\
95$^{c}$    &   3394745453589896832 &  0.40  &     0.24$\pm$0.12  &  172.1$\pm$14.2  &  0.83$\pm$0.15  &  126.0$\pm$ 5.3  &  0.55$\pm$0.03  &  123.8$\pm$ 1.8  &  0.63$\pm$0.05  &  103.0$\pm$ 2.2  \\
96*$^{b}$    &   3394745453589894272 &  0.46  &     0.57$\pm$0.17  &  121.9$\pm$ 8.8  &  0.77$\pm$0.13  &  126.7$\pm$ 4.8  &  0.98$\pm$0.10  &  113.5$\pm$ 2.9  &  0.53$\pm$0.15  &  124.4$\pm$ 7.9  \\
97*$^{a}$    &   3394745655451733760 &  0.38  &     1.30$\pm$0.20  &  161.2$\pm$ 4.5  &  1.05$\pm$0.05  &  128.4$\pm$ 1.5  &  0.86$\pm$0.09  &  125.8$\pm$ 3.1  &  0.44$\pm$0.07  &  130.4$\pm$ 4.3  \\
98    &   3394745659748306176 &  0.31  &     0.52$\pm$0.26  &  112.9$\pm$14.4  &  	   -          &       -     &  0.61$\pm$0.23  &  121.3$\pm$10.8  &  0.45$\pm$0.05  &  142.7$\pm$ 3.0  \\
99*$^{a}$    &   3394744663316152192 &  0.38  &     0.38$\pm$0.10  &  119.3$\pm$ 7.6  &  0.71$\pm$0.01  &  110.3$\pm$ 0.5  &  0.86$\pm$0.11  &  106.4$\pm$ 3.7  &  0.36$\pm$0.15  &   99.9$\pm$12.2  \\
100   &   3394745797187276416 &  0.33  &     1.03$\pm$0.16  &  131.0$\pm$ 4.5  &  1.42$\pm$0.11  &  114.9$\pm$ 2.3  &  0.97$\pm$0.11  &  120.6$\pm$ 3.2  &  0.83$\pm$0.09  &  111.7$\pm$ 3.2  \\
101$^{b}$   &   3394744693379350912 &  0.28  &     	 -          &  	 -	       &  0.73$\pm$0.02  &   99.7$\pm$ 0.8  &  0.95$\pm$0.01  &  117.6$\pm$ 0.5  &  0.53$\pm$0.03  &  110.4$\pm$ 1.9  \\
102*$^{a}$   &   3394745694108039040 &  0.28  &     0.93$\pm$0.26  &  103.6$\pm$ 8.1  &  0.90$\pm$0.05  &   99.7$\pm$ 1.6  &  0.65$\pm$0.16  &   98.8$\pm$ 6.9  &      -    	&     - 	   \\
103   &   3394756311267231488 &  0.42  &           -	    &  	 -	       &  0.97$\pm$0.19  &  125.1$\pm$ 5.6  &  0.85$\pm$0.13  &  118.5$\pm$ 4.5  &  0.68$\pm$0.12  &  124.2$\pm$ 5.3  \\
104   &   3394743426365371776 &  0.16  &     0.56$\pm$0.09  &  115.9$\pm$ 4.5  &  0.91$\pm$0.14  &  114.6$\pm$ 4.4   &  0.65$\pm$0.23  &  109.7$\pm$10.2  &      -      	&     - 	   \\
105*   &   3394743117127933312 &  0.32  &     0.86$\pm$0.39  &  117.1$\pm$13.1  &  1.31$\pm$0.16  &  116.9$\pm$ 3.4   &  2.91$\pm$0.92  &   86.3$\pm$ 9.2  &      -    	&     - 	   \\
106   &   3394736313899758208 &  0.42  &     0.56$\pm$0.08  &  179.5$\pm$ 4.3  &  0.76$\pm$0.13  &  116.7$\pm$ 4.7   &  0.74$\pm$0.17  &  113.7$\pm$ 6.4  &  0.16$\pm$0.07  &  118.1$\pm$12.5  \\
107*$^{a}$   &   3394742429932982272 &  0.25  &     0.75$\pm$0.26  &  149.5$\pm$ 9.8  &  0.64$\pm$0.24  &  113.6$\pm$10.5   &  0.51$\pm$0.04  &  117.7$\pm$ 2.4  &  0.20$\pm$0.08  &  145.3$\pm$11.0  \\
108*$^{a}$   &   3394742429932979712 &  0.25  &     0.28$\pm$0.12  &  166.1$\pm$12.4  &  0.96$\pm$0.11  &  112.1$\pm$ 3.3   &  0.86$\pm$0.10  &  118.2$\pm$ 3.4  &  0.42$\pm$0.01  &  118.4$\pm$ 0.2  \\
109*$^{a}$   &   3394742429933179648 &  0.25  &     0.42$\pm$0.11  &  106.9$\pm$ 7.9  &  0.66$\pm$0.22  &  107.4$\pm$ 9.4   &  0.65$\pm$0.02  &  120.9$\pm$ 1.0  &  0.35$\pm$0.06  &  112.7$\pm$ 5.3  \\
110   &   3394736348259494656 &  0.26  &     0.62$\pm$0.23  &   65.6$\pm$10.6  &  1.16$\pm$0.03  &   89.8$\pm$ 0.8   &  0.46$\pm$0.19  &  141.4$\pm$11.9  &  0.45$\pm$0.08  &  107.0$\pm$ 5.0  \\
111   &   3394733187163568128 &  0.27  &     0.44$\pm$0.21  &  176.1$\pm$13.9  &  	  -           &      -       &  0.82$\pm$0.17  &  104.0$\pm$ 5.9  &  0.29$\pm$0.07  &  104.0$\pm$ 7.1  \\
112*   &   3394742155055273728 &  0.24  &     0.48$\pm$0.22  &  142.2$\pm$13.1  &  0.84$\pm$0.20  &   83.3$\pm$ 6.7   &      -	       &  	  -	      &      -    	&    -  	   \\
113   &   3394739200117777152 &  0.37  &     0.55$\pm$0.12  &  141.9$\pm$ 6.4  &  0.85$\pm$0.18  &  133.7$\pm$ 6.1   &      -	       &  	  -	      &      -    	&    -  	   \\
114   &   3394742292494221952 &  0.25  &     0.23$\pm$0.05  &  100.7$\pm$ 6.2  &  0.45$\pm$0.22  &  142.4$\pm$14.1   &      -	       &  	  -	      &      -    	&    -  	   \\
115   &   3394742704811075456 &  0.87  &    0.57$\pm$0.19  &  137.0$\pm$ 9.4   &  0.95$\pm$0.24  &  108.8$\pm$ 7.4   &      -	       &  	  -	      &      -    	&    -  	   \\
116$^{f}$   &   3394742739170812800 &  0.43  &    0.84$\pm$0.13  &  103.3$\pm$ 4.5   &  0.96$\pm$0.06  &   97.0$\pm$ 1.9   &      -	       &  	  -	      &      -    	&    -  	   \\
117*$^{b}$   &   3394739303196980224 &  0.19  &    0.47$\pm$0.11  &  101.5$\pm$ 6.6   &  0.60$\pm$0.05  &  115.9$\pm$ 2.6   &      -	       &  	  -	      &      -    	&    -  	   \\
118   &   3394736691856720512 &  1.03  &       -	   &  	 -	       &     -	       	 &  	 - 	     &  0.47$\pm$0.21  &  115.5$\pm$13.0  &  0.36$\pm$0.08  &  124.3$\pm$ 6.7  \\
119*   &   3394737207252771840 &  0.74  & 	-      	   &  	 -	       &     -	       	 &  	 - 	     &  0.95$\pm$0.11  &   86.8$\pm$ 3.2  &  0.70$\pm$0.29  &  102.0$\pm$12.0  \\	
120   &   3394737443474461440 &  0.37  &  	-      	   &  	 -	       &     -	      	 &  	 - 	     &  0.83$\pm$0.04  &  112.6$\pm$ 1.4  &  0.36$\pm$0.05  &  100.5$\pm$ 3.7  \\
121   &   3394736485698262912 &  0.58  &  	-      	   &  	 -	       &     -	      	 &  	 - 	     &  0.64$\pm$0.11  &  117.2$\pm$ 5.0  &  0.33$\pm$0.16  &  118.1$\pm$13.7  \\
122*$^{b}$   &   3394735798503688704 &  0.37  &  	 -	   &  	 -	       &     -	      	 &  	 - 	     &  1.25$\pm$0.07  &  112.9$\pm$ 1.6  &  0.72$\pm$0.07  &   79.2$\pm$ 2.6  \\ 
123   &   3394737379051441920 &  0.58  &  	 -	   &  	 -  	       &     -	      	 &  	 - 	     &  0.69$\pm$0.08  &  103.1$\pm$ 3.4  &  0.44$\pm$0.02  &  117.1$\pm$ 1.6  \\ 
124*$^{a}$   &   3394743357645905792 &  0.59  &  	  -	   &  	 -  	       &     -	      	 &  	 - 	     &  0.83$\pm$0.11  &  100.9$\pm$ 3.6  &  0.51$\pm$0.01  &  119.2$\pm$ 0.1  \\		
125*   &   3394743288926431360 &  0.55  &  	  -	   &  	 -  	       &     -	      	 &  	 - 	     &  0.55$\pm$0.03  &  133.8$\pm$ 1.8  &  0.34$\pm$0.03  &  112.3$\pm$ 2.6  \\
 \enddata
 \end{deluxetable*}
\tablecomments{* Member stars, $^{a}$ giants \citep{2020AJ....159...96S},  $^{b}$ pulsating stars \citep{2020AJ....159...96S}, $^{c}$ blue stragglers \citep{2007A&A...463..789A}, $^{d}$ eclipsing binaries \citep{2020AJ....159...96S}, $^{e}$ blue horizontal branch stars \citep{2010A&A...522A..88S}, and $^{f}$ pre-main-sequence stars \citep{2018A&A...620A.172Z}.  \\ $P_{B}$, $P_{V}$, $P_{R}$, and $P_{I}$ are the degrees of polarization in the B, V, R, and I bands, respectively. $\theta_{B}$, $\theta_{V}$, $\theta_{R}$, and $\theta_{I}$ are the polarization angles in the B, V, R, and I bands,  respectively.}
 \end{longrotatetable}

\begin{table*}
    \centering
    \caption{Average Values of $P$ and $\theta$ in the B, V, R, and I Bands For Members, Nonmembers and All Sources in the Region of the Cluster NGC 1817.}\label{tab:ptheta}
    \begin{tabular}{c|ccccc}
    \hline
    \hline
               &\multicolumn{1}{c}{Members}     && \multicolumn{1}{c}{NonMembers}  && \multicolumn{1}{c}{All}\\
         \hline
         \hline
$P_{B}$ &     0.75$\pm$0.02    &&   0.77$\pm$0.04      &&   0.75$\pm$0.02 \\
$\theta_{B}$ & 133.3$\pm$0.9   &&   133.7$\pm$1.3      &&   133.5$\pm$0.8 \\
\hline
$P_{V}$   &  0.91$\pm$0.02        &&   1.01$\pm$0.03  &&   0.95$\pm$0.02\\
$\theta_{V}$ &    118.6$\pm$0.6   &&  120.0$\pm$0.8   &&    119.2$\pm$0.5 \\
\hline
$P_{R}$     &  0.90$\pm$0.02        &&    0.84$\pm$0.03   &&  0.88$\pm$0.02\\
$\theta_{R}$  &  115.1$\pm$0.6      &&    120.6$\pm$1.0   &&  117.5$\pm$0.6 \\
\hline
$P_{I}$      & 0.55$\pm$0.02        &&   0.46$\pm$0.02   &&   0.51$\pm$0.01 \\
$\theta_{I}$   & 119.5$\pm$1.0      &&   120.3$\pm$1.3   &&   119.8$\pm$0.8 \\
\hline
\hline
    \end{tabular}
   \end{table*}

\startlongtable
\begin{deluxetable*}{lccr|lccr}
\tablecolumns{6}
\vspace{-0.1in}	
\centering
\tablecaption{Parameters $P_{max}$, $\lambda_{max}$, and $\sigma_{1}$ as Obtained from Fitting the Serkowski Relation.	\label{tab:serkowski}}
\tablewidth{20pt}
\setlength\tabcolsep{20.0pt}
\tablehead{
\colhead{ID} & \colhead{$P_{max}$} & \colhead{$\lambda_{max}$} & \colhead{$\sigma_{1}$} & \colhead{ID} & \colhead{$P_{max}$} & \colhead{$\lambda_{max}$} &  \colhead{$\sigma_{1}$} \\
\nocolhead{~~} & \colhead{(\%)} & \colhead{($\mu m$)} & \nocolhead{~~} & \nocolhead{~~} &\colhead{(\%)} &  \colhead{($\mu m$)}}
\startdata
 1*    &  1.05$\pm$0.02   &   0.56$\pm$0.03   &   0.3    &    54*   &  3.11$\pm$0.44   &   0.20$\pm$0.01   &   0.1  \\
2*    &  0.98$\pm$0.05   &   0.65$\pm$0.04   &   0.7    &     56    &  0.71$\pm$0.13   &   0.55$\pm$0.16   &   0.8  \\
3     &  0.98$\pm$0.34   &   0.87$\pm$0.36   &   1.2    &     58    &  1.84$\pm$0.63   &   0.20$\pm$0.03   &   0.3  \\
4*    &  1.28$\pm$0.38   &   0.31$\pm$0.05   &   0.6    &     60*   &  0.96$\pm$0.12   &   0.69$\pm$0.16   &   0.5  \\
5*    &  0.83$\pm$0.11   &   0.40$\pm$0.05   &   0.6    &     61*   &  0.85$\pm$0.19   &   0.52$\pm$0.22   &   1.1  \\
6     &  1.28$\pm$0.40   &   0.39$\pm$0.12   &   0.4    &     64*   &  1.61$\pm$0.65   &   0.33$\pm$0.11   &   1.0  \\
7     &  0.75$\pm$0.02   &   0.56$\pm$0.05   &   0.6    &     65*   &  2.44$\pm$1.04   &   0.27$\pm$0.07   &   0.7  \\
9*    &  1.24$\pm$0.56   &   0.29$\pm$0.10   &   0.8    &     66*   &  1.24$\pm$0.29   &   0.34$\pm$0.06   &   0.4  \\
10*   &  0.87$\pm$0.34   &   0.34$\pm$0.10   &   0.6    &     67    &  0.78$\pm$0.10   &   0.49$\pm$0.09   &   0.3  \\
11*   &  0.69$\pm$0.04   &   0.58$\pm$0.09   &   0.2    &     68*   &  1.61$\pm$0.06   &   0.30$\pm$0.01   &   0.4  \\
12*   &  0.89$\pm$0.01   &   0.55$\pm$0.07   &   0.5    &     69    &  0.82$\pm$0.25   &   0.81$\pm$0.30   &   1.0  \\
13*   &  0.71$\pm$0.10   &   0.56$\pm$0.15   &   1.7    &     70    &  1.88$\pm$0.73   &   0.25$\pm$0.04   &   0.6  \\
15*   &  0.80$\pm$0.10   &   0.52$\pm$0.11   &   0.5    &     74    &  1.15$\pm$0.38   &   0.96$\pm$0.20   &   1.5  \\
16*   &  1.05$\pm$0.18   &   0.36$\pm$0.05   &   0.4    &     75*   &  0.73$\pm$0.11   &   0.67$\pm$0.11   &   0.7  \\
17*   &  1.45$\pm$0.08   &   0.39$\pm$0.04   &   0.4    &     76    &  0.83$\pm$0.10   &   0.76$\pm$0.27   &   0.8  \\
18    &  4.83$\pm$1.23   &   0.21$\pm$0.02   &   0.2    &     79    &  0.76$\pm$0.02   &   0.40$\pm$0.02   &   0.2  \\
19*   &  1.09$\pm$0.06   &   0.55$\pm$0.10   &   0.6    &     80    &  1.36$\pm$0.59   &   0.35$\pm$0.10   &   0.9  \\
23*   &  0.77$\pm$0.07   &   0.50$\pm$0.08   &   0.4    &     81*   &  0.91$\pm$0.15   &   0.93$\pm$0.13   &   0.6  \\
24*   &  1.18$\pm$0.17   &   0.34$\pm$0.04   &   1.1    &     85*   &  0.97$\pm$0.19   &   0.51$\pm$0.17   &   1.2  \\
25*   &  0.90$\pm$0.02   &   0.61$\pm$0.07   &   0.2    &     86*   &  0.92$\pm$0.03   &   0.53$\pm$0.06   &   0.8  \\
26*   &  0.68$\pm$0.04   &   0.54$\pm$0.04   &   0.2    &     87*   &  0.91$\pm$0.05   &   0.49$\pm$0.05   &   0.3  \\
27*   &  1.28$\pm$0.38   &   0.72$\pm$0.33   &   2.4    &     89*   &  1.41$\pm$0.02   &   0.64$\pm$0.02   &   0.1  \\
28    &  0.82$\pm$0.16   &   0.43$\pm$0.11   &   0.5    &     90*   &  0.80$\pm$0.09   &   0.43$\pm$0.07   &   0.5  \\
29*   &  1.07$\pm$0.20   &   0.45$\pm$0.10   &   0.2    &     91    &  0.44$\pm$0.12   &   0.83$\pm$0.29   &   1.1  \\
32*   &  1.15$\pm$0.06   &   0.35$\pm$0.01   &   0.1    &     92*   &  0.85$\pm$0.06   &   0.45$\pm$0.04   &   0.4  \\
35    &  1.60$\pm$0.01   &   0.48$\pm$0.01   &   0.0    &     93*   &  0.94$\pm$0.18   &   0.43$\pm$0.06   &   0.9  \\
38    &  0.77$\pm$0.02   &   0.52$\pm$0.06   &   0.1    &     94*   &  0.61$\pm$0.08   &   0.57$\pm$0.10   &   0.5  \\
40*   &  1.41$\pm$0.46   &   0.41$\pm$0.14   &   2.2    &     95    &  0.63$\pm$0.14   &   0.92$\pm$0.31   &   0.7  \\
41*   &  0.82$\pm$0.12   &   0.48$\pm$0.20   &   1.2    &     96*   &  0.82$\pm$0.12   &   0.65$\pm$0.22   &   0.6  \\
42*   &  0.99$\pm$0.07   &   1.03$\pm$0.04   &   0.1    &     97*   &  1.71$\pm$0.41   &   0.29$\pm$0.04   &   0.4  \\
43    &  1.46$\pm$0.03   &   0.49$\pm$0.03   &   0.1    &     98    &  0.57$\pm$0.08   &   0.51$\pm$0.07   &   0.2  \\
44    &  1.36$\pm$0.05   &   0.43$\pm$0.02   &   0.4    &     99*   &  0.74$\pm$0.15   &   0.68$\pm$0.27   &   0.9  \\
45*   &  0.76$\pm$0.05   &   0.62$\pm$0.08   &   0.5    &     100   &  1.26$\pm$0.22   &   0.45$\pm$0.09   &   0.6  \\
46*   &  0.82$\pm$0.02   &   0.39$\pm$0.01   &   0.1    &     102*  &  1.07$\pm$0.20   &   0.37$\pm$0.07   &   0.3  \\
47    &  1.08$\pm$0.36   &   0.40$\pm$0.18   &   1.0    &     103   &  1.04$\pm$0.02   &   0.44$\pm$0.01   &   0.0  \\
48*   &  0.61$\pm$0.02   &   0.54$\pm$0.07   &   0.2    &     104   &  0.87$\pm$0.40   &   0.78$\pm$0.34   &   0.6  \\
51*   &  1.09$\pm$0.25   &   0.72$\pm$0.26   &   0.9    &     109*  &  0.67$\pm$0.17   &   0.52$\pm$0.22   &   0.8  \\
53    &  1.32$\pm$0.20   &   0.36$\pm$0.06   &   0.3    &   \\
\enddata
\end{deluxetable*}

\newpage
\begin{table*}
\centering
 \caption{Average Values of $P_V$ and $\theta_V$ for the Different Types of Stars in the Cluster NGC 1817.} \label{tab:var_avg}
\begin{tabular}{lcccc}
\hline
Type           &       $P_{V}$  &    $\theta_{V}$     \\
\hline
Giants                  & 0.89$\pm$0.02     &  118.5$\pm$0.9   \\
Blue Stragglers         & 0.87$\pm$0.04     &  124.1$\pm$1.4 \\
Pulsating$^{\star}$     & 0.85$\pm$0.03     &  117.9$\pm$1.1  \\
~~~~~~~~~$\delta$ Scuti & 0.86$\pm$0.04     &  118.3$\pm$1.6   \\
~~~~~~~~~$\gamma$ Dor   & 0.83$\pm$0.03     &  112.5$\pm$1.1   \\
~~~~~~~~~hybrid         & 0.87$\pm$0.07     &  119.5$\pm$2.4  \\
Eclipsing               & 0.87$\pm$0.03     &  118.2$\pm$1.3   \\
Members$^{\star\star}$  & 0.96$\pm$0.04     &  117.3$\pm$1.2   \\

\hline
\end{tabular}
~\\
$^{\star}$ Average value $P_V$ and $\theta_V$ for the pulsating stars $\delta$ Scuti, $\gamma$ Dor, and hybrid pulsators.\\
$^{\star\star}$ Average values $P_V$ and $\theta_V$ for members stars after excluding variable stars.
\end{table*}

\end{document}